\documentclass[conference]{IEEEtran}
\IEEEoverridecommandlockouts

\usepackage{cite}
\usepackage{amsmath,amssymb,amsfonts}
\usepackage{graphicx}
\usepackage{textcomp}
\usepackage{xcolor}
\usepackage{algorithm}
\usepackage{algpseudocode}
\usepackage{subcaption}
\usepackage{tabularx}
\usepackage{comment}
\usepackage{caption}
\usepackage{makecell}
\usepackage{booktabs}
\usepackage{cite} 
\usepackage[colorlinks=true, linkcolor=blue, citecolor=blue, urlcolor=blue]{hyperref}

\usepackage{libertine}
\usepackage[libertine]{newtxmath} 
\usepackage{setspace}
\AtBeginDocument{%
  \providecommand\BibTeX{{%
    Bib\TeX}}}
\usepackage{verbatim}
\usepackage{algorithm}
\usepackage{algpseudocode}
\usepackage{subcaption}
\usepackage{array}
\usepackage[scaled=0.85]{inconsolata}
\usepackage{fancyhdr}
\let\oldalgorithm\algorithm
\renewcommand{\algorithm}{
    \oldalgorithm
    \footnotesize
}
\def\BibTeX{{\rm B\kern-.05em{\sc i\kern-.025em b}\kern-.08em
    T\kern-.1667em\lower.7ex\hbox{E}\kern-.125emX}}
\begin{document}

\title{GPU-Accelerated Parallel Selected Inversion for Structured Matrices Using \emph{sTiles}}
\author{}

\author{
Esmail Abdul Fattah, Hatem Ltaief, H{\aa}vard Rue, and David Keyes\\
Division of Computer, Electrical, and Mathematical Sciences and Engineering (CEMSE),\\
King Abdullah University of Science and Technology (KAUST),\\
Thuwal, 23955, Makkah, Saudi Arabia\\
\{esmail.abdulfattah, hatem.ltaief, haavard.rue, david.keyes\}@kaust.edu.sa
}


\maketitle

\begin{abstract}
    Selected inversion is essential for applications such as Bayesian inference, electronic structure calculations, and inverse covariance estimation, where computing only specific elements of large sparse matrix inverses significantly reduces computational and memory overhead. We present an efficient implementation of a two-phase parallel algorithm for computing selected elements of the inverse of a sparse symmetric matrix \( A \), which can be expressed as \( A = LL^T \) through sparse Cholesky factorization. Our approach leverages a tile-based structure, focusing on selected dense tiles to optimize computational efficiency and parallelism. While the focus is on arrowhead matrices, the method can be extended to handle general structured matrices. Performance evaluations on a dual-socket 26-core Intel Xeon CPU server demonstrate that \emph{sTiles}\footnote{https://github.com/esmail-abdulfattah/sTiles} outperforms state-of-the-art direct solvers such as Panua-PARDISO, achieving up to 13X speedup on large-scale structured matrices. Additionally, our GPU implementation using an NVIDIA A100 GPU demonstrates substantial acceleration over its CPU counterpart, achieving up to 5X speedup for large, high-bandwidth matrices with high computational intensity. These results underscore the robustness and versatility of \emph{sTiles}, validating its effectiveness across various densities and problem configurations.
\end{abstract}

\section{Introduction}

Matrix inversion is a fundamental operation in numerical linear algebra, which is pivotal to numerous applications in science and engineering. However, inverting large, sparse symmetric matrices is computationally intensive, particularly when dealing with specialized structures like arrowhead matrices. These matrices, characterized by non-zero elements concentrated along the block diagonal, the last block row, and the last block column, are prevalent in fields such as mathematics, physics, and engineering.

Traditional inversion methods often compute the entire inverse matrix, leading to a loss of sparsity as originally sparse structures are transformed into dense ones. This phenomenon can be interpreted through the lens of the Cayley-Hamilton theorem, which states that every square matrix satisfies its characteristic equation. As a result, when inverting matrices symbolically or computationally, terms in the matrix power series contribute to non-zero elements in all positions, resulting in the proliferation of fill-ins. Consequently, this highlights the inefficiency of direct inversion approaches for sparse matrices and motivates the need for tailored techniques that preserve sparsity.
\begin{figure}
    \centering
    \includegraphics[width=0.35\linewidth]{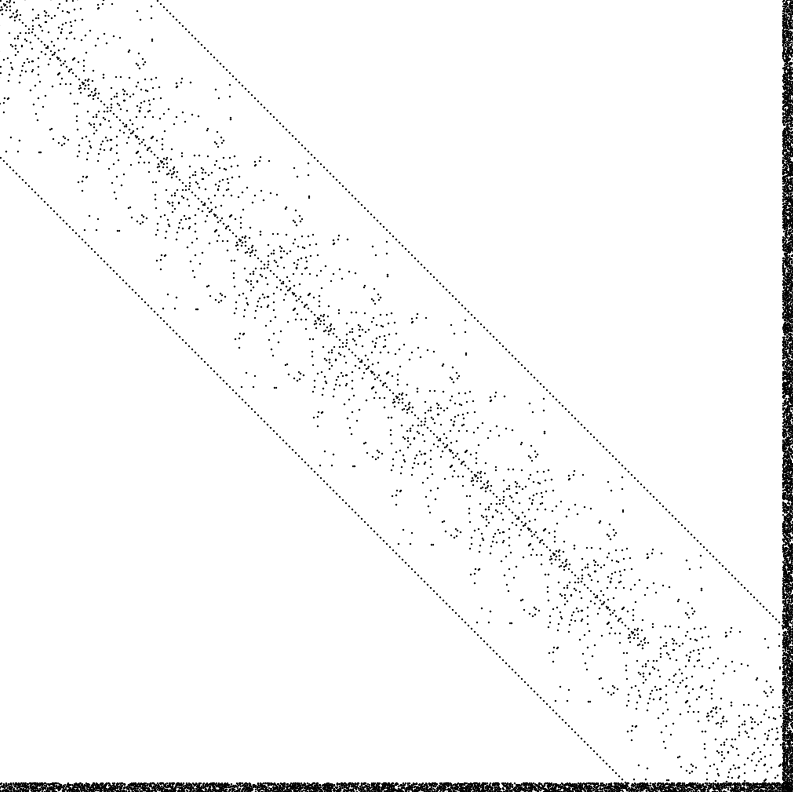}
    \hspace{0.05\linewidth}
    \includegraphics[width=0.35\linewidth]{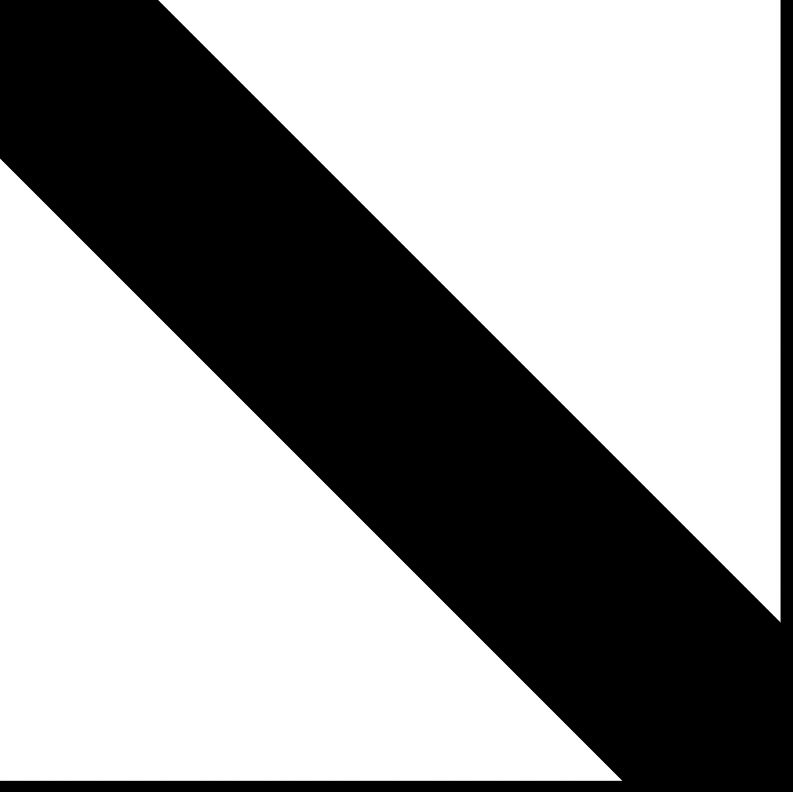}

    \vspace{1em} 

    \includegraphics[width=0.35\linewidth]{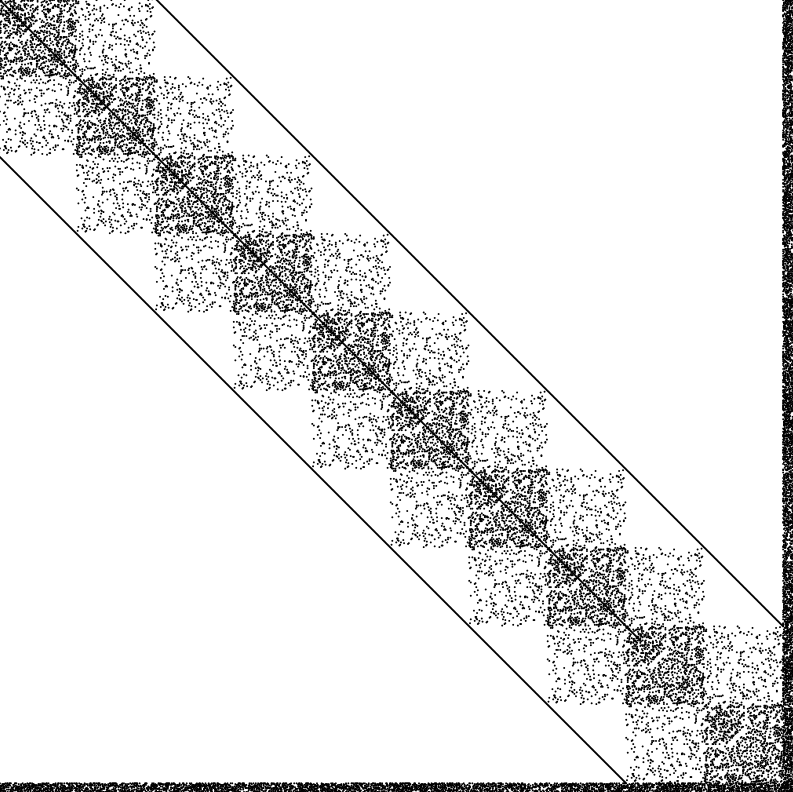}
    \hspace{0.05\linewidth}
    \includegraphics[width=0.35\linewidth]{smat3.png}

    \caption{Top row: matrix \textit{A} (left) and the selected inverse of matrix \textit{A} (right). Bottom row: matrix \textit{B} (left) and the selected inverse of matrix \textit{B} (right).}
    \label{fig:illutssparsity}
\end{figure}
To address this challenge, selected inversion techniques have emerged as a focused alternative, computing only specific entries or substructures of the inverse matrix that are required for the target application. By avoiding the computation of elements considered unnecessary, the selected inversion significantly reduces resource consumption. This makes it particularly valuable in domains such as large-scale sparse inverse covariance estimation \cite{bollhofer2019large}, electronic structure calculations \cite{lin2009multipole}, and Bayesian modeling \cite{zhumekenov2023parallel}. However, despite these advantages, achieving scalability for high-dimensional problems in selected inversion methods remains a significant challenge.

The sparsity patterns of arrowhead matrices offer unique computational advantages, particularly in Bayesian inference, where they model interactions between multiple random effects. These matrices are commonly constructed using Kronecker products, for example, by capturing space-time dependencies in spatio-temporal models \cite{lindgren2024diffusion}. Figure~\ref{fig:illutssparsity} illustrates this by comparing two arrowhead matrices (matrix \textit{A} and matrix \textit{B}) with differing sparsity levels. The sparsity level of these matrices often reflects the number of spatial locations or time points used in a model. As data collection technologies advance, we are entering an era of increasingly high-resolution datasets, where more sensors, satellites, and monitoring systems generate vast amounts of spatial and temporal information, leading to lower levels of sparsity in arrowhead matrices, driven by higher resolution and wider matrix bandwidth, as the number of locations increases.

Arrowhead matrices of this form are widely used in various scientific domains, particularly in Bayesian modeling \cite{rue2009approximate}, including geosciences, where they are employed to model climate variations and assess geological risks \cite{fioravanti2023interpolating}. In epidemiology, they play a crucial role in tracking the spread of infectious diseases over time and space \cite{myer2019spatiotemporal}, and have been adopted in public health efforts such as excess mortality estimation by the World Health Organization (WHO) \cite{who2022excess} and analyses of teen birth rates and drug poisoning mortality by the Centers for Disease Control and Prevention (CDC) \cite{cdc2024teenbirths}. Biostatistics benefits from these matrices in patient health modeling, particularly for analyzing longitudinally and spatially distributed data \cite{moraga2019geospatial}. They are also integral to environmental modeling (e.g., predicting air pollution levels and ecological trends \cite{seaton2024spatio} and global demographic studies such as the United Nations' World Population Prospects 2024 \cite{un2024wpp}). Additionally, meteorology relies on them to model weather patterns based on historical spatio-temporal data.

Despite their structured sparsity, computing the full inverse of such matrices results in a fully dense matrix, leading to significant memory and computational overhead. However, by selectively computing only the inverse of the non-zero elements in matrices \textit{A} and \textit{B}, we preserve their arrowhead structure, reducing storage and computational costs. As illustrated in Figure~\ref{fig:illutssparsity}, regardless of the initial sparsity level, the selected inverse retains a dense arrowhead shape. This highlights the need of working with fine-grained data structures (e.g., tiles) from the outset, as it enables optimized computation and memory management without unnecessarily transforming the problem into a fully dense form.

To address the challenges of scalability and hardware efficiency, we build upon the tile-based paradigm. Tile-based methods provide finer granularity, enhance data locality, and expose parallelism for superior scalability, making them especially effective for structured matrices like arrowheads. \emph{sTiles} \cite{fattah2025stiles}, a tile-based framework for high-performance linear algebra, exemplifies these advantages by utilizing sparse-dense tile computations for efficient factorizations and subsequent matrix operations. By design, \emph{sTiles} minimizes inter-task dependencies and optimizes parallel execution, forming a robust foundation for extending selected inversion techniques. In this work, we extend the functionality of \emph{sTiles} to efficiently perform selected inversion, significantly broadening its practical applicability to broader matrix structures.

The remainder of this article is organized as follows. Section~\ref{sec:related_work} reviews related work in selected inversion. Section~\ref{sec:parallel_algo} introduces our parallel tile-based algorithm, beginning with its conceptual foundation in recursive inversion and detailing the two-phase implementation for both CPUs and GPUs. Section~\ref{sec:performance} presents a comprehensive performance evaluation, including comparisons with state-of-the-art libraries and an analysis of GPU acceleration. Finally, Section~\ref{sec:conclusion} concludes with a summary of our findings and discusses potential future directions.

\section{Related Work}
\label{sec:related_work}

Our work uses a direct, factorization-based method for selected inversion, building upon the foundation laid by Takahashi's formula \cite{takahashi1973formation}. Modern high-performance solvers in this domain, such as MUMPS \cite{amestoy2001fully} and PSelInv \cite{jacquelin2016pselinv}, typically employ supernodal or multifrontal techniques. These methods group columns with similar sparsity patterns into ``supernodes" to leverage high-performance BLAS-3 kernels, but their irregular data structures can be challenging to optimize for massively parallel hardware.

Among state-of-the-art solvers for shared-memory systems, Panua-PARDISO \cite{schenk2001pardiso} is a highly optimized library that uses a robust multi-level parallelization scheme. A comprehensive analysis by Verbosio \cite{verbosio2019high} demonstrated that Panua-PARDISO is significantly faster and more memory-efficient than competitors like PSelInv for the class of sparse-dense problems relevant to our work. These findings establish Panua-PARDISO as the state-of-the-art and our primary benchmark.

In contrast to the supernodal approach, our algorithm is tile-based. We impose a uniform grid of fixed-size blocks (tiles) onto the matrix, creating a highly regular data structure and a predictable dependency pattern. This regularity allows us to employ a static scheduling strategy that maximizes data locality and minimizes runtime overhead, making our approach exceptionally well-suited for structured matrices and hybrid CPU-GPU architectures.

\section{Parallel Selected Inversion} \label{sec:parallel_algo}

Our selected inversion algorithm is built upon the Cholesky factorization of a symmetric positive-definite matrix, \(A = LL^T\). The inverse \(\Sigma = A^{-1}\) is computed by solving the block-wise matrix equation \(L^T \Sigma = L^{-1}\) recursively. This tile-based approach decomposes the problem into a sequence of high-performance operations (e.g., \textbf{TRSM}, \textbf{LAUUM}, \textbf{GEMM}, and \textbf{TRMM}) on small, dense tiles, which enhances data locality and exposes fine-grained parallelism.

\subsection{Recursive Tile-Based Inversion via Cholesky Decomposition} \label{sec:recursive_inversion}

Efficient inversion of large symmetric positive-definite matrices is central to many scientific applications, particularly when only selected entries of the inverse are needed. By leveraging the Cholesky factorization and operating at the tile granularity, we enable an approach that exploits data locality, promotes parallelism, and scales well across hardware platforms.

Cholesky decomposition provides a natural foundation for this strategy. Given a matrix 
\textit{A}, the decomposition
\[
A = LL^T,
\]
\noindent where \(L\) is a lower triangular matrix, forms the basis for many matrix inversion algorithms. Traditional approaches compute the inverse of \(A\) by inverting \(L\) element-wise and then computing \((L^{-1})^T\), but these methods are often inefficient for large matrices. To overcome these limitations, tile-based algorithms decompose \(L\) into smaller blocks or tiles, enabling parallel computation of the inverse while managing interdependencies between tiles to maintain accuracy.

Early work by Agullo et al.\ introduced a tile-based in-place algorithm for the full inversion of symmetric positive-definite matrices, leveraging dynamic scheduling and compiler-inspired techniques such as loop reversal and array renaming to improve parallelism on multicore architectures \cite{agullo2011towards}. This laid the foundation for asynchronous, task-based inversion strategies in dense linear algebra.

To illustrate, the derivation of the full inversion is based on the relationship
\[
L^T \Sigma = L^{-1},
\]
\noindent where both \(L\) and \(\Sigma\) are represented in a tile-based structure. For simplicity, consider \(3 \times 3\) tiles. The transpose of the matrix \(L\) is given as:
\[
L^T = \begin{bmatrix}
L_{00}^T & L_{10}^T & L_{20}^T \\
0 & L_{11}^T & L_{21}^T \\
0 & 0 & L_{22}^T
\end{bmatrix},
\quad
\Sigma = \begin{bmatrix}
\Sigma_{00} & \Sigma_{01} & \Sigma_{02} \\
\Sigma_{10} & \Sigma_{11} & \Sigma_{12} \\
\Sigma_{20} & \Sigma_{21} & \Sigma_{22}
\end{bmatrix},
\]
\vspace{0.5em}

\noindent where \(\Sigma_{ij}\) and \(L_{ij}\) are the submatrices corresponding to the \((i,j)\)-th tile in the respective matrices. Each tile can be either dense or sparse, depending on the matrix structure and sparsity pattern.

We start the recursive computation with the tile \(\Sigma_{22}\), using the relationship:
\[
L_{22}^T \Sigma_{22} = L_{22}^{-1}, \quad \Sigma_{22} = L_{22}^{-T} L_{22}^{-1}.
\]
Next, we compute \(\Sigma_{21}\) as follows:
\[
L_{11}^T \Sigma_{21}^T + L_{21}^T \Sigma_{22} = 0, \quad \Sigma_{21} = -\Sigma_{22} L_{21} L_{11}^{-1}.
\]

These computations propagate recursively to compute other tiles of \(\Sigma\). The relationships for each tile are summarized below:
\[
\begin{array}{rl}
\Sigma_{22} & = L_{22}^{-T} L_{22}^{-1}, \\[6pt]
\Sigma_{21} & = -\Sigma_{22} L_{21} L_{11}^{-1}, \\[6pt]
\Sigma_{11} & = - L_{11}^{-T} L_{21}^T \Sigma_{21} + L_{11}^{-T} L_{11}^{-1}, \\[6pt]
\Sigma_{20} & = -\Sigma_{21} L_{10} L_{00}^{-1} - \Sigma_{22} L_{20} L_{00}^{-1}, \\[6pt]
\Sigma_{10} & = -\Sigma_{11} L_{10} L_{00}^{-1} - \Sigma_{12} L_{20} L_{00}^{-1}, \\[6pt]
\Sigma_{00} & = - L_{00}^{-T} L_{10}^{T} \Sigma_{10} -  L_{00}^{-T} L_{20}^T \Sigma_{20} + L_{00}^{-T} L_{00}^{-1}.
\end{array}
\]
\vspace{0.5em}

The equations demonstrate how the inverse tiles are computed recursively, starting from the bottom-right corner of \(\Sigma\) and propagating through the dependencies to the top-left corner. Algorithm \ref{alg:serial_phase} outlines the tile-based inversion procedure for a full matrix \( A \) using its Cholesky decomposition \( A = LL^T \).

The algorithm starts by computing diagonal tiles \(\Sigma_{ii}\) using the relationship \(\Sigma_{ii} = L_{ii}^{-T} L_{ii}^{-1}\), followed by updating off-diagonal tiles \(\Sigma_{ji}\) through recursive propagation of dependencies. The key operations involve matrix multiplications and additions performed at the tile level, ensuring efficient computation.

\begin{algorithm}
\caption{Tile-based inversion of a full matrix} \label{alg:serial_phase}
\begin{algorithmic}[1]
\State \textbf{Initialization:}
\State \texttt{int i, j, k;}
\For{$i = N-1$ \textbf{to} $0$ \textbf{step} $-1$}
    \For{$j = N-1$ \textbf{to} $i$ \textbf{step} $-1$}
        \If{$i == j$}
            \State $\Sigma_{ii} \gets \Sigma_{ii} + L_{ii}^{-T} L_{ii}^{-1}$
            \For{$k = i+1$ \textbf{to} $N$}
                \State $\Sigma_{ii} \gets \Sigma_{ii} - \Sigma_{ik} L_{ki} L_{ii}^{-1}$
            \EndFor
        \Else
            \State $\Sigma_{ji} \gets \Sigma_{ji} - \Sigma_{jj} L_{ji} L_{ii}^{-1}$
            \For{$k = i+1$ \textbf{to} $N$}
                \If{$j != k$}
                    \State $\Sigma_{ji} \gets \Sigma_{ji} - \Sigma_{jk} L_{ki} L_{ii}^{-1}$
                \EndIf
            \EndFor
        \EndIf
    \EndFor
\EndFor
\end{algorithmic}
\end{algorithm}

The core tile operations required for recursive inversion are implemented using well-established matrix computational kernels. These operations are as follows:

\begin{itemize}
    
    \item \textbf{TRSM (Triangular Solve with Multiple Right-Hand Sides)}: This kernel computes the inverse of a diagonal tile by solving a triangular system with the identity matrix:
    \[
    \Sigma_{ii} \gets \Sigma_{ii}^{-1}.
    \]

    \item \textbf{LAUUM (Lower Triangular Matrix Multiplication)}: This kernel updates a diagonal tile. It takes the lower triangular part of the tile (denoted as $L_{ii}$) and computes its product with its transpose. The algorithm then mirrors the result to make the tile fully symmetric.
    \[
    \Sigma_{ii} \gets L_{ii} L_{ii}^T.
    \]

    \item \textbf{GEMM (General Matrix-Matrix Multiplication)}: This operation updates an off-diagonal tile by performing matrix multiplication and subtracting the result:
    \[
    \Sigma_{ji} \gets \Sigma_{ji} - \Sigma_{kj}^T \Sigma_{ki}.
    \]

    \item \textbf{TRMM (Triangular Matrix-Matrix Multiplication)}: This kernel updates off-diagonal tiles by multiplying them with a triangular matrix:
    \[
    \Sigma_{ji} \gets L_{jj} \Sigma_{ji}.
    \]
\end{itemize}

In this work, we build on the tile paradigm but focus on structured sparse matrices, extending the tile-based framework to \textit{selected inversion}, where only specific entries of the inverse are computed. This selective approach introduces significant computational savings by avoiding unnecessary operations: if a given operation (e.g., computing a particular tile update) does not contribute to any of the user-requested tiles, it is skipped entirely. This stands in contrast to the full inversion approach outlined in Algorithm~\ref{alg:serial_phase}, where all tiles are processed regardless of necessity. In our implementation, we adapt this algorithm by incorporating a filtering mechanism that prunes irrelevant computations while preserving correctness.

\subsection{Selected Elements of the Inverse}

Building upon the tile-based sparse Cholesky factorization implemented in \emph{sTiles}~\cite{fattah2025stiles}, our selected inversion algorithm adapts this concept by pruning all computations not required for the user-specified inverse elements. The overall process is structured into the following key steps:

\begin{enumerate}
    \item The \textit{selection step} begins with the user specifying a list of matrix elements, identified by their indices \((i, j)\), for which the inverse is required. These indices are mapped to their corresponding tiles using the same compressed tile format employed during the Cholesky decomposition.
    \item A \textit{symbolic inversion} step follows, during which all necessary dependencies for the selected elements are identified and incorporated. This step guarantees the completeness and accuracy of the subsequent computations.
    \item Finally, the \textit{numerical inversion} is performed, computing the desired inverse elements. This step leverages the tile-based structure to maximize computational efficiency.
\end{enumerate}

\begin{figure}
    \centering
    \rotatebox{90}{
        \includegraphics[height=0.45\textwidth]{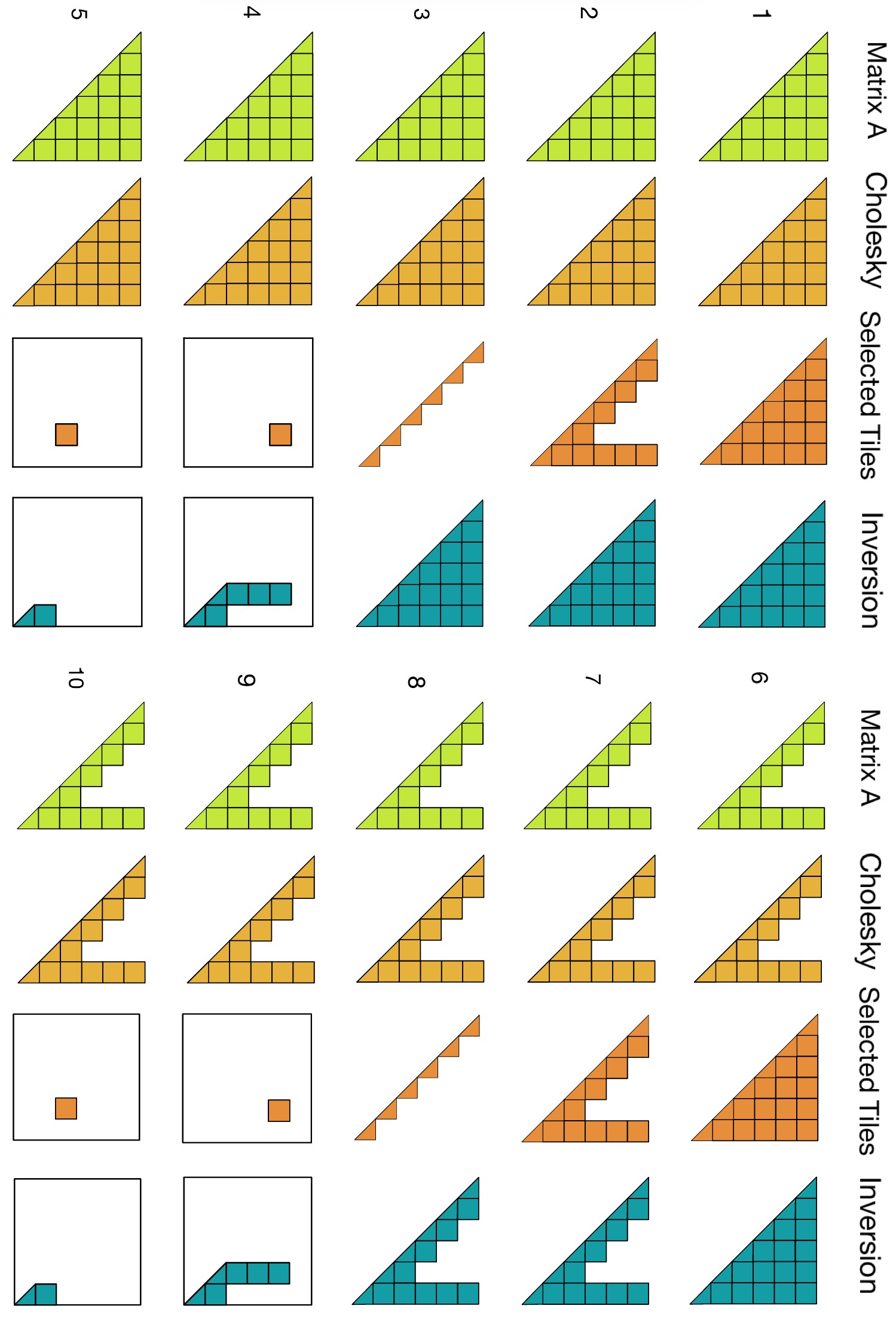}
    }
    \caption{Illustration of Cholesky factorization and inversion patterns of selected tiles for dense (1-5) and arrowhead (6-10) symmetric matrices.}
    \label{fig:selinvillus}
\end{figure}

Next, we delve into the implications of this process by examining different matrix structures and their corresponding inversion patterns, with a particular focus on dense and arrowhead matrices.

Consider the task of computing the inverse for a selected set of element pairs within a matrix. The matrix is partitioned into tiles, and as an initial step, these elements are mapped to their corresponding tiles, referred to as \textit{selected tiles}. Although this approach necessitates computing the inverse for all elements within a tile, even if only a subset is of primary interest, it provides significant advantages at negligible cost in memory and computational complexity. This computation is essential due to fill-in, as many elements within these tiles become populated during the inversion process. Additionally, from a memory efficiency standpoint, using tiles improves cache utilization, thereby enhancing overall computational performance.

Our focus is on computing the selected inverse of structured matrices, with particular attention to arrowhead patterns. Figure~\ref{fig:selinvillus} presents ten illustrative cases divided into two groups: fully dense matrices (cases 1-5) and arrowhead matrices (cases 6-10). For each case, we show the original matrix \( A \), its Cholesky factor, the selected tiles requested for inversion, and the resulting inversion pattern.

In cases 1-3, the original matrix is fully dense, and the selected tiles include the diagonal. As a result, the inversion covers the entire lower triangle, producing a full inverse. This behavior reflects the high computational cost of inverting dense matrices when the diagonal is involved. Case 6 mirrors this behavior in the arrowhead setting. Although the original structure is sparse, selecting the entire matrix leads to full inversion, analogous to case 1.

Cases 7 and 8 represent arrowhead matrices where the selected tiles form an arrowhead structure that includes the diagonal. In these cases, the resulting inverse retains the same arrowhead structure. This behavior closely matches that of cases 2 and 3 in the dense setting, where the selected tiles include the diagonal and do not extend beyond the original non-zero pattern.

Cases 4-5 and 9-10 show a consistent pattern across both dense and arrowhead matrices: the selected tiles do not include any diagonal tiles. Consequently, only a minimal subset of the inverse is computed, and the cost remains low. These cases highlight how omitting the diagonal and restricting selection to isolated tiles significantly reduces the computational effort required for inversion.

Our focus is on the arrowhead matrix in which the selected pattern matches the Cholesky pattern, specifically case 7. This formulation can be adapted to case 6 when needed. We next present the Directed Acyclic Graph (DAG) representing the inversion process for cases 2 and 7.

\subsection{Directed Acyclic Graph}

The Directed Acyclic Graph (DAG) captures dependencies between computations, enabling efficient parallel execution and optimized resource allocation. By organizing tasks with well-defined precedence, the DAG ensures that independent computations can proceed concurrently. The DAG underlying our approach is constructed using four key computational kernels—\textbf{TRSM}, \textbf{LAUUM}, \textbf{GEMM}, and \textbf{TRMM}—each applied to specific tiles \(\Sigma_{i,j}\), corresponding to the tile at position \((i,j)\). The definitions and roles of these operations are described in detail in Section~\ref{sec:recursive_inversion}.

\begin{figure*}[t]
    \centering
    \includegraphics[width=0.8\textwidth]{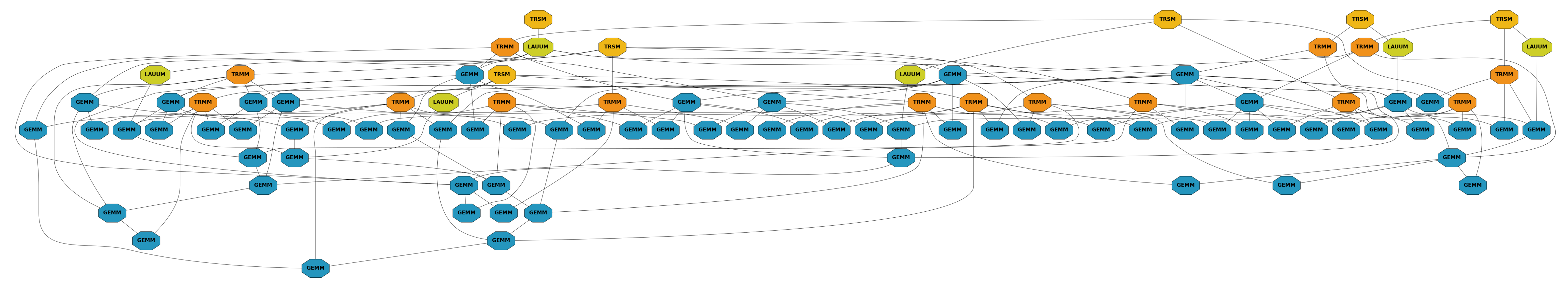}
    \caption{Directed acyclic graph (DAG) for selected inversion on a dense matrix of size \(6 \times 6\) tiles, corresponding to Case 2 in Figure~\ref{fig:selinvillus}. The DAG's width illustrates the high degree of parallelism available. Its height represents the length of the critical path, which dictates the minimum execution time.}
    \label{fig:dag_full}
\end{figure*}

\begin{figure*}[t]
    \centering
    \includegraphics[width=0.6\textwidth]{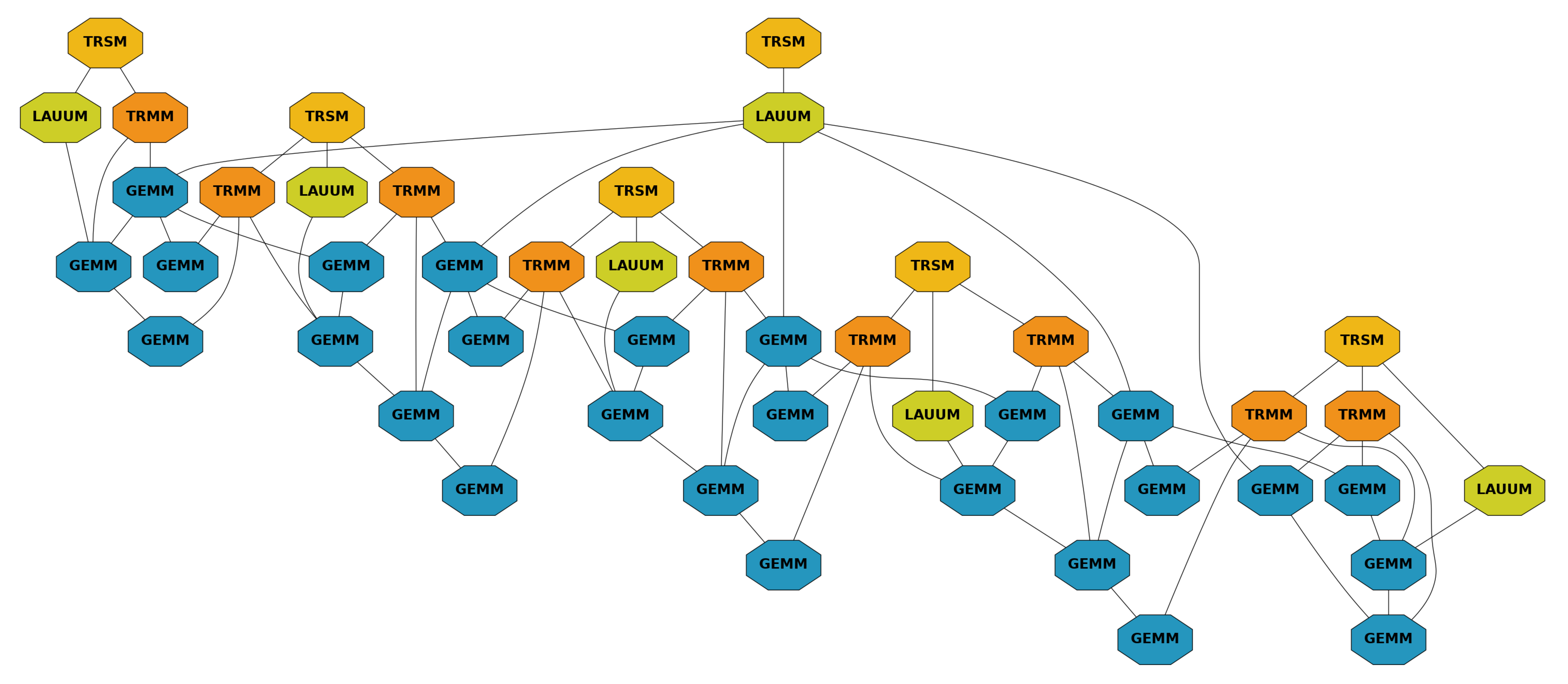}
    \caption{Directed acyclic graph (DAG) for selected inversion on an arrowhead matrix of size \(6 \times 6\) tiles, corresponding to Case 7 in Figure~\ref{fig:selinvillus}. Compared to the dense case in Figure~\ref{fig:dag_full}, this DAG has significantly fewer nodes, demonstrating the computational savings from exploiting the matrix structure, while the critical path length remains the same.}
    \label{fig:dag_arrowhead}
\end{figure*}

As illustrated in Figures~\ref{fig:dag_full} and \ref{fig:dag_arrowhead}, the DAGs for full and arrowhead matrix inversions (Cases 2 and 7, respectively) exhibit notable differences in structure and parallelism. The DAG for full matrix inversion is characterized by a large width, indicating many tasks that could, in principle, be executed concurrently. In the case of the arrowhead structure, the number of nodes is significantly reduced. This visual difference is a direct result of the filtering mechanism mentioned in Section~\ref{sec:recursive_inversion}, which prunes unnecessary computations by evaluating only the dependencies required for the user-selected tiles. This leads to a more compact DAG with a reduced overall computational workload. Crucially, the critical path length is unaffected, remaining at six sequential operations along the longest dependency chain for the studied matrices in both the full and arrowhead cases.

\subsection{Two-Phase Algorithm for Selected Inversion}

To parallelize Algorithm~\ref{alg:serial_phase}, we adopt a two-phase approach designed to minimize interdependencies between cores, thereby reducing idle times and enhancing parallel efficiency. This approach serves as the practical implementation of the filtering mechanism, as it prunes unnecessary computations by structuring its loops to operate only on tiles that contribute to the user-specified subset of the inverse. By leveraging the structure of the selected tiles, we avoid redundant operations while preserving correctness. The division of the algorithm into distinct phases enables better task organization and facilitates scalable parallel execution. A critical aspect of this strategy is the use of static load balancing, where tasks are preassigned to cores during preprocessing. This ensures an even distribution of the workload and significantly reduces runtime scheduling overhead.\\

\noindent \textbf{Phase 1: Independent Row-wise Updates of \(L^T\)}

In this phase, we begin the inversion of the upper triangular factor \(L^T\) by performing a set of computationally independent row-wise operations. The parallelization is achieved by distributing the tile rows of \(L^T\) among the available cores. The tile rows are assigned in a static, round-robin manner: a core with a given \texttt{thread ID} is responsible for processing the set of block rows indexed by \(i\), starting from \(i = N - 1 - \texttt{thread ID}\) and decrementing by \texttt{total\_cores} in each step.

This distribution scheme is highly efficient as the computations for different tile rows are completely independent of one another. For any given tile row \(i\), all operations only require data from tiles within that same row (i.e., \(L_{ii}^T\) and \(L_{ij}^T\)). Therefore, cores can proceed concurrently without any need for communication or synchronization during this phase.

The work for each assigned row \(i\) consists of two steps, as detailed in Algorithm~\ref{alg:phase1}:
\begin{enumerate}
    \item \textbf{Diagonal Tile Inversion:} The diagonal tile \(L_{ii}^T\) is inverted. The result, \((L_{ii}^T)^{-1}\), is stored in the corresponding diagonal block of the output matrix, \(\Sigma_{ii}^T\). This corresponds to the \texttt{TRSM} operation in the algorithm.
    \item \textbf{Off-Diagonal Tile Update:} The newly computed \((L_{ii}^T)^{-1}\) is then used to update all non-zero off-diagonal tiles \(L_{ij}^T\) (for \(j > i\)) in the same block row. This update, performed by the \texttt{TRMM} operation, modifies the \(L^T\) matrix in-place, preparing it for the second phase of the algorithm.
\end{enumerate}

To handle sparse matrices, we use the notation \( j \in \text{neighbors}(i) \) to indicate that an operation is only performed if the tile is structurally non-zero. A non-zero tile may have an initial value of zero but is filled during the symbolic factorization phase and is fully updated by the end of the computation. The algorithm for this phase is detailed in Algorithm~\ref{alg:phase1}. 

\begin{algorithm}
\caption{Parallel selected inversion - phase 1} \label{alg:phase1}
\begin{algorithmic}[1]
\State \textbf{Given:} Matrix $A$ is partitioned into $N \times N$ tiles and expressed as $A = LL^T$, where $L^T$ is the upper triangular factor.
\State \textbf{Note:} $\Sigma$ is the matrix where the selected inverse is stored.
\State \textbf{Initialization:} $i \gets N - 1 - \texttt{thread ID}$
\While{$i \geq 0$}
        \For{\textbf{all} $j \in \text{neighbors}(i)$ \textbf{and} $i \leq  j < N - 1$}
            \If{$i == j$}
            \State $\Sigma_{ii}^T \gets \texttt{TRSM}(L_{ii}^T, I)$ 
            \Else
                \State $L_{ij}^T \gets \texttt{TRMM}(L_{ii}^T, L_{ij}^T)$ 
            \EndIf
        \EndFor-
    \State $i \gets i - \texttt{total\_cores}$
\EndWhile
\end{algorithmic}
\end{algorithm}

\noindent \textbf{Phase 2: Dependent Tile Computations}

Once Phase 1 is complete, the algorithm transitions to a second phase to finalize the selected inverse, $\Sigma$. Unlike the first phase, this stage involves complex data dependencies between tile computations. The calculation of a tile $\Sigma_{ij}$ often requires results from other tiles (e.g., $\Sigma_{kj}$), which may be computed by different cores. These dependencies create a task graph where cores must synchronize to ensure correct computational ordering.
To manage these dependencies, we implement a lightweight, asynchronous producer-consumer model. This model is realized using a shared status-tracking matrix, \texttt{core\_progress}, which allows threads to signal the completion of a tile and wait for dependencies without expensive global barriers.
The producer-consumer mechanism works as follows:
\begin{itemize}
    \item \textbf{Producer Role:} When a core successfully computes a tile $\Sigma_{xy}$, it acts as a producer by updating the corresponding entry in \texttt{core\_progress} to a "completed" state. This update signals to all other cores that the data in $\Sigma_{xy}$ is now ready to be used.
    \item \textbf{Consumer Role:} Before a core computes a tile that depends on $\Sigma_{xy}$, it acts as a consumer. It polls the status of the prerequisite tile by reading the \texttt{core\_progress} matrix, waiting until it is marked as complete.
\end{itemize}
For clarity in the pseudocode (Algorithm~\ref{alg:phase2}), we abstract this mechanism into two high-level primitives:
\begin{itemize}
\item \texttt{WaitForTile(row, col)}: Represents the consumer's action of waiting for the flag in core-progress to be set.
\item \texttt{SignalTileReady(row, col)}: Represents the producer's action of setting the flag upon task completion.
\end{itemize}
This approach allows the algorithm to focus on the logical dataflow, enabling a high degree of parallelism as cores can work on any available tasks whose dependencies have been met.

\begin{algorithm}
\caption{Parallel selected inversion - phase 2} \label{alg:phase2}
\begin{algorithmic}[1]
\State \textbf{Define synchronization primitives:}
\State \texttt{WaitForTile(row, col)}: Pauses until tile $\Sigma_{\texttt{row, col}}$ is marked complete.
\State \texttt{SignalTileReady(row, col)}: Marks tile $\Sigma_{\texttt{row, col}}$ as complete.

\Statex
\State \textbf{Initialization:} $i \gets N - 1 -$ thread ID and set = false;
\While{$i \geq 0$}
    \For{$j = N - 1$ \textbf{to} $i$ \textbf{step} $-1$}
        \If{$i == j$} 
            \State $\Sigma_{ii} \gets \texttt{LAUUM}(\Sigma_{ii})$
            \For{\textbf{all} $k \in \text{neighbors}(i)$ \textbf{and} $i < k < N - 1$} 
                \State \texttt{WaitForTile(i, k)}
                \State $\Sigma_{ii} \gets \texttt{GEMM}(L_{ik}, \Sigma_{ik}^T, \Sigma_{ii})$
            \EndFor
            \State \texttt{SignalTileReady(i,i)} 
        \Else
            \If{$(i+1) \leq (N - 1)$} 
                \State \texttt{set = true; ii = 0; jj = 0;}
            \EndIf
            \For{\textbf{all} $k \in \text{neighbors}(j)$ \textbf{and} $i < k < N - 1$}
                \If{$i \in \text{neighbors}(j) \cap j \in \text{neighbors}(k)$ \textbf{and} $i < j$}
                    \If{$k > j$}
                        \State \texttt{WaitForTile(j, k)}
                        \State $\Sigma_{ij} \gets \texttt{GEMM}(L_{ik}, \Sigma_{kj}, \Sigma_{ij})$
                        \State \texttt{ii = i; jj = j;}
                    \Else
                        \State \texttt{ WaitForTile(k, j)}
                        \State $\Sigma_{ij} \gets \texttt{GEMM}(L_{ik}, \Sigma_{jk}^T, \Sigma_{ij})$
                        \State \texttt{ii = i; jj = j;}
                    \EndIf
                \EndIf
            \EndFor
            \If{\texttt{set}} 
                \State \texttt{SignalTileReady(ii, jj)}
            \EndIf
        \EndIf
    \EndFor
    \State $i \gets i - \texttt{total\_cores}$
\EndWhile
\end{algorithmic}
\end{algorithm}

\begin{figure*}[t]
    \centering
    \begin{subfigure}[b]{\linewidth}
        \centering
        \includegraphics[width=0.8\linewidth]{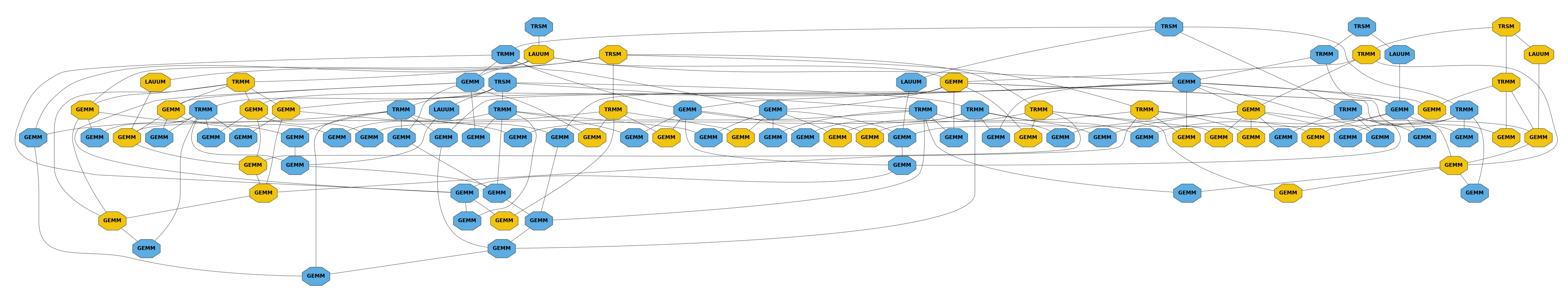}
        \caption{Full matrix inversion using 2 cores.}
        \label{fig:dag_full_2cores}
    \end{subfigure}
    
    \vspace{1em} 
    
    \begin{subfigure}[b]{\linewidth}
        \centering
        \includegraphics[width=0.8\linewidth]{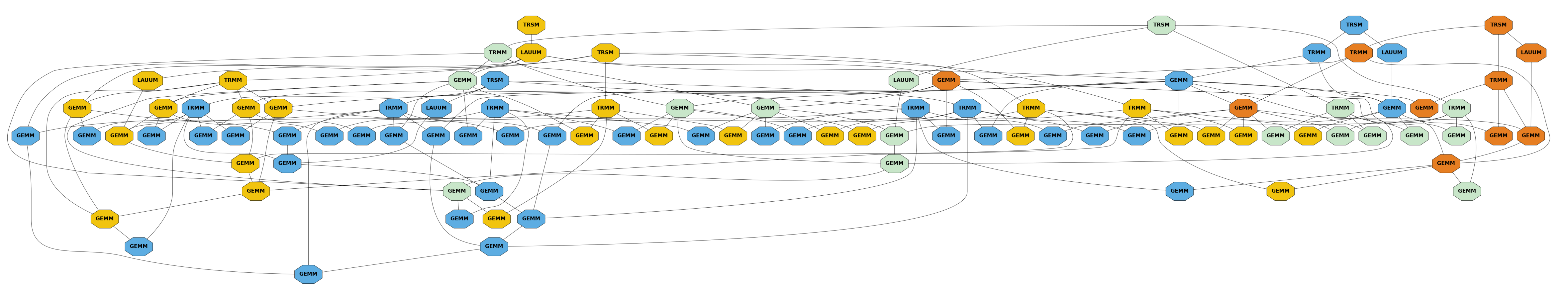}
        \caption{Full matrix inversion using 4 cores.}
        \label{fig:dag_full_4cores}
    \end{subfigure}
    
    \vspace{1em}
    
    \begin{subfigure}[b]{0.48\linewidth}
        \centering
        \includegraphics[width=\linewidth]{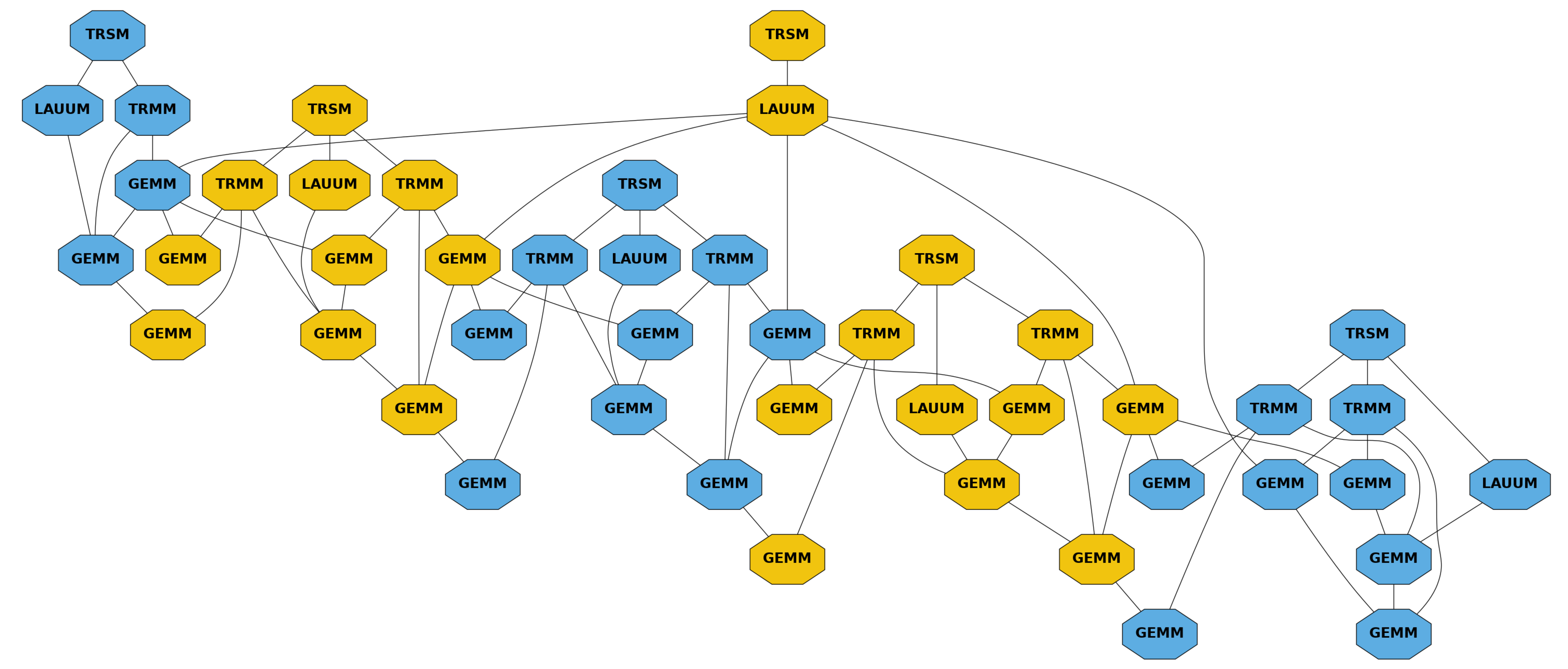}
        \caption{Arrowhead matrix inversion using 2 cores.}
        \label{fig:dag_arr_2cores}
    \end{subfigure}
    \hfill 
    \begin{subfigure}[b]{0.48\linewidth}
        \centering
        \includegraphics[width=\linewidth]{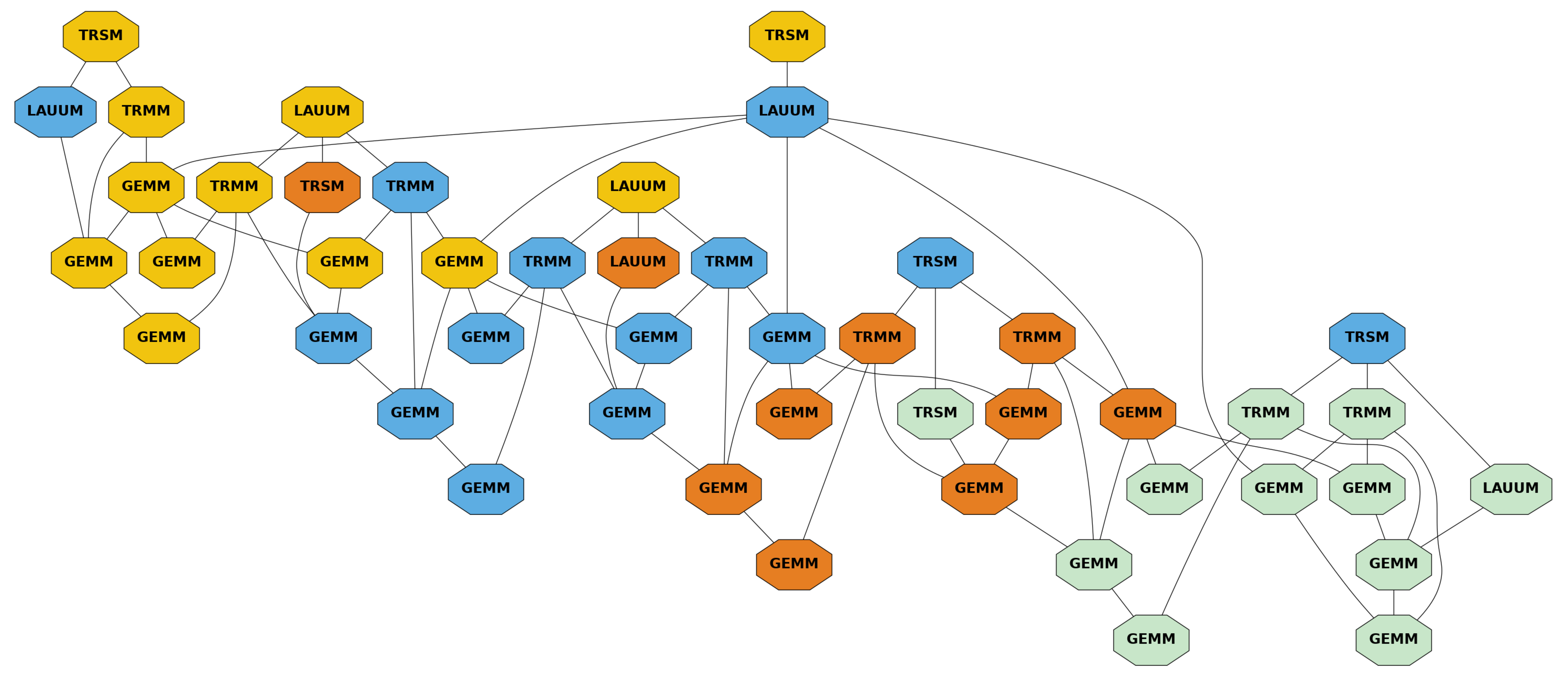}
        \caption{Arrowhead matrix inversion using 4 cores.}
        \label{fig:dag_arr_4cores}
    \end{subfigure}

    \caption{Directed acyclic graphs (DAGs) illustrating the task distribution for selected inversion. The top two rows show the DAGs for a \textbf{full matrix} with 2 and 4 cores, respectively; their width illustrates the high degree of potential parallelism. The bottom row shows the significantly more compact DAGs for a structured \textbf{arrowhead matrix}, highlighting the drastic reduction in the number of tasks. Each color represents tasks assigned to a specific core.}
    \label{fig:dagsarrowfullcores}
\end{figure*}

The Directed Acyclic Graphs (DAGs) in Figure~\ref{fig:dagsarrowfullcores} directly represent the task distribution and parallelism strategies implemented in the two-phase algorithm detailed above. The graphs clearly demonstrate how tasks are assigned to different cores, with each color representing the tasks handled by a specific core. While this visualization uses a matrix of size 6x6 tiles for illustrative clarity, the principles of static task distribution and parallelism it demonstrates are fundamental to the algorithm's scalability for larger problems. For larger matrices, this structured partitioning leads to a significantly greater number of concurrent tasks, fully exploiting the available computational resources.

\subsection{Algorithmic Complexity}
\label{sec:complexity_main}

A matrix of size \(n \times n\) is tiled into \(N \times N\) blocks of size \(b \times b\), so \(n = N\,b\). The cost of a single tile operation, such as \textbf{GEMM} or \textbf{TRSM}, is \(O(b^{3})\) floating-point operations.

For a full, dense matrix, the complexity of our block inversion algorithm is dominated by \textbf{GEMM} operations, resulting in a total workload of \(O(n^3)\), which is consistent with standard methods.

In contrast, for the selected inversion of an arrowhead matrix with a band of width consisting of \(B\) tile blocks, the workload is significantly reduced. By only performing computations that contribute to the selected arrowhead pattern, the complexity becomes:
\[
W_{\text{selected}} = O(B^2 n b^2)
\]
This complexity is asymptotically lower than the dense \(O(n^3)\) case, making the method highly efficient when the band \(B\) is small relative to the number of tiles \(N\). Table~\ref{tab:complexity_comparison} illustrates this drastic reduction in the number of required block operations. A detailed derivation of these complexity results is provided in Appendix~\ref{app:complexity_derivation}.

\begin{table}
  \centering
  \caption{Comparison of block operation counts for full dense vs. selected inversion.}
  \label{tab:complexity_comparison}
  \footnotesize
  \setlength{\tabcolsep}{3pt} 
  \begin{tabular*}{\columnwidth}{@{\extracolsep{\fill}} l c c }
    \toprule
    \textbf{Inversion} & \textbf{GEMM Ops} & \textbf{Asymptotic} \\
    \textbf{Type}      & \textbf{(\(N=6\))} & \textbf{Trend}      \\
    \midrule
    Full (\(B=6\)) & 70 & \(O(N^3) \to O(n^3)\) \\
    \midrule
    Selected (\(B=1\)) & 10 & \(O(N) \to O(n b^2)\) \\
    Selected (\(B=2\)) & 26 & \(O(N) \to O(n b^2)\) \\
    \bottomrule
  \end{tabular*}
\end{table}

\subsection{GPU Acceleration for Parallel Selected Inversion}

The GPU implementation of parallel selected inversion in \textit{sTiles} follows a similar tile-based approach as its CPU counterpart, with adaptations to leverage the massive parallelism and high throughput of modern GPUs. Given that selected inversion focuses only on specific elements of the inverse, we assume that the selected tiles can fit entirely within a single GPU's memory, ensuring efficient execution without the need for frequent data transfers between the CPU host and the GPU device.

The same tile size criteria used in the GPU implementation of Cholesky factorization in \textit{sTiles} is adopted for the GPU version of the selected inversion to maintain consistency in memory access patterns and workload distribution. Each tile is mapped to a dedicated CUDA stream, enabling concurrent execution of independent tasks across multiple GPU compute units.

To optimize performance and minimize data transfer overhead, the entire matrix, along with its factorization, is fully copied to GPU memory before any computations begin. All computational kernels in the CPU implementation, such as Cholesky factorization, triangular solves, symmetric rank-k updates, and matrix multiplications, are replaced with their respective cuBLAS and cuSOLVER implementations. The key operations include: \\
\indent - Triangular solve: \texttt{cublasDtrsm} \\ 
\indent - Triangular matrix multiply: \texttt{cublasDtrmm} \\ 
\indent - Matrix multiplication: \texttt{cublasDgemm}

Once all computations are completed, the results are transferred back to the CPU for further processing or storage. Since data movement between CPU and GPU is a major bottleneck, the design ensures that all required computations are performed on the GPU before any data is transferred back, maximizing throughput and reducing unnecessary communication overhead.

This GPU-accelerated implementation of parallel selected inversion in \textit{sTiles} significantly improves performance for structured matrices by fully utilizing GPU resources, minimizing data transfers, and leveraging parallel execution through CUDA streams.

While our current implementation assumes that the selected tiles fit in GPU memory, the algorithm could be extended to support out-of-core execution for larger matrices that exceed GPU capacity. This would involve overlapping data movement and computation while maintaining efficient static scheduling. Such strategies have been successfully applied in the context of out-of-core Cholesky factorization on modern GPU architectures, as demonstrated in \cite{ren2024accelerating}. Incorporating similar techniques into \textit{sTiles} would enable selected inversion to scale to larger problem sizes on next-generation heterogeneous systems such as the NVIDIA Grace Hopper Superchip with a unified memory subsystem on the CPU host.

\section{Performance Evaluation and Experimental Results} \label{sec:performance}

\begin{figure}
    \centering
    \includegraphics[width=0.4\columnwidth]{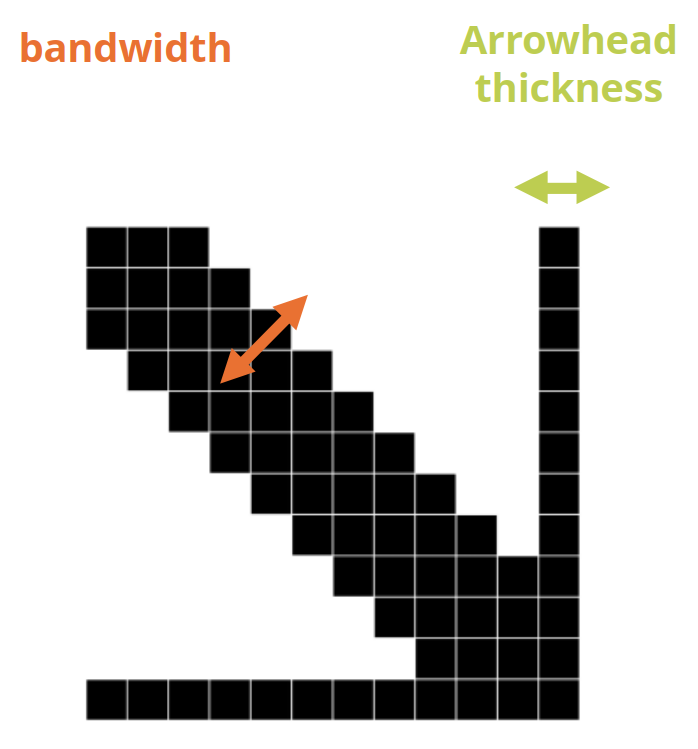}
    \caption{Visual illustration of matrix structure showing bandwidth and arrowhead thickness. These dimensions influence the density and shape of the matrices used in our performance evaluation. In typical Bayesian inference modeling such as the INLA method \cite{rue2009approximate}, the arrowhead thickness corresponds to the number of fixed effects included in the model, while the bandwidth reflects the local structure of random effects.}
    \label{fig:matrix_structure_illustration}
\end{figure}

In this section, we present a comprehensive evaluation of our GPU-accelerated parallel selected inversion implementation using \textit{sTiles}. While we compare its CPU implementation against an existing state-of-the-art CPU-only approach, our goal for the GPU implementation is to demonstrate the impact of the \textit{sTiles} fine-grained algorithmic approach on massively parallel hardware accelerators.

\vspace{-2pt}
\subsection{Experimental Setup and Software} \label{setup}

Our computational experiments were conducted on two high-performance computing (HPC) systems, each used for a specific set of evaluations involving \textit{sTiles}:

\vspace{-4pt}
\begin{itemize}
    \item \textbf{CPU Server}: A dual-socket 26-core system featuring Intel\textsuperscript{\textregistered} Xeon\textsuperscript{\textregistered} Gold 6230R processors, with 52 cores total operating at 2.10 GHz and a total L3 cache of 71.5 MB. This system was used for all CPU-only experiments, including scalability, density sensitivity, and full matrix inversion. Additionally, the Panua-PARDISO license is available on this system, enabling direct solver comparisons under a fully licensed environment.
    
    \item \textbf{GPU Node}: A dedicated compute node incorporating 64 AMD EPYC 7713 CPU cores running at 1.99 GHz, complemented by a single NVIDIA A100-SXM4 GPU operating at 1.16 GHz and equipped with 80 GB of high-bandwidth memory (HBM2). This system was used for all GPU-accelerated experiments.
\end{itemize}

\vspace{-3pt}
In comparisons between CPU and GPU performance (e.g., Table~\ref{tab:cpu_gpu_comparison}), the CPU baseline was measured using the AMD EPYC 7713 CPU on the GPU node to ensure consistency in hardware and avoid bias introduced by architectural differences.

To benchmark our parallel selected inversion algorithm, we compare it against \textbf{Panua-PARDISO 8.2}, a state-of-the-art direct solver for sparse linear systems utilizing $LL^T$ factorization, optimized for shared-memory parallelism. This solver employs advanced numerical techniques to ensure efficient factorization and inversion, making it a robust reference for our comparative analysis. To isolate the impact of the selected inversion algorithm, we compare only the \textbf{selected inversion time} in our performance evaluation. The Cholesky factorization time is excluded from these comparisons. A detailed analysis of the Cholesky phase has already been presented separately in \cite{fattah2025stiles}.

\subsection{Performance Evaluation}

\begin{table}
  \caption{Matrix properties used in Cholesky factorization and selected inversion experiments for \textit{sTiles} -- Set 1. These matrices reflect the arrowhead structures that commonly arise in INLA-based models \cite{rue2009approximate}.}
  \label{tab:matrix_properties_stiles1}
  \centering
  \scriptsize
  \setlength{\tabcolsep}{4pt} 
  \begin{tabular*}{\columnwidth}{@{\extracolsep{\fill}} c c c c c}
    \toprule
    \makecell{\textbf{Matrix}\\\textbf{ID}} & \textbf{Size} & \textbf{Bandwidth} & \makecell{\textbf{Arrowhead}\\\textbf{Thickness}} & \makecell{\textbf{Density}\\\textbf{(\%)}} \\
    \midrule
    1  & 10,010   & 100   & 10  & 0.408  \\
    2  & 10,010   & 200   & 10  & 0.605  \\
    3  & 10,010   & 300   & 10  & 0.643  \\
    4  & 10,200   & 100   & 200 & 3.938  \\
    5  & 10,200   & 200   & 200 & 4.032  \\
    6  & 10,200   & 300   & 200 & 4.066  \\
    7  & 100,010  & 1000  & 10  & 0.121  \\
    8  & 100,010  & 2000  & 10  & 0.219  \\
    9  & 100,010  & 3000  & 10  & 0.258  \\
    10 & 100,200  & 1000  & 200 & 0.498  \\
    11 & 100,200  & 2000  & 200 & 0.597  \\
    12 & 100,200  & 3000  & 200 & 0.637  \\
    13 & 500,010  & 1000  & 10  & 0.024  \\
    14 & 500,010  & 2000  & 10  & 0.044  \\
    15 & 500,010  & 3000  & 10  & 0.052  \\
    16 & 500,200  & 1000  & 200 & 0.100  \\
    17 & 500,200  & 2000  & 200 & 0.120  \\
    18 & 500,200  & 3000  & 200 & 0.128  \\
    \bottomrule
  \end{tabular*}
\end{table}

Table~\ref{tab:matrix_properties_stiles1} summarizes the structured matrices utilized in our evaluation, encompassing variations in size, bandwidth, and sparsity levels. These matrices are carefully chosen to represent practical applications, such as statistical models with structured sparsity patterns. Specifically, the \textit{arrowhead thickness} corresponds to the number of fixed effects in statistical models, with values ranging from 10 (moderate case) to 200 (extreme case), see Figure \ref{fig:matrix_structure_illustration}. The benchmarking aims to assess the computational efficiency of \textit{sTiles} in handling these structures and to compare its performance against Panua-PARDISO 8.2, a state-of-the-art solver.

Our benchmarking methodology consists of the following steps:

\begin{enumerate}
    \item \textbf{Factorization and Selected Inversion:} Cholesky factorization is performed, followed by selected inversion using \textit{sTiles} and Panua-PARDISO 8.2. For each specific test case (a given matrix on a given number of cores), a single execution was performed to measure the runtime of the selected inverse.

    \item \textbf{Parallel Scalability Evaluation:} To assess parallel performance, these execution times were recorded across a wide range of core counts (1, 2, 4, 8, 16, 32, and 52 cores, the maximum available on our test system). This allows us to observe how each solver's performance scales with increasing parallelism.

    \item \textbf{Performance Analysis and Visualization:} The results are analyzed in terms of absolute runtime and relative speedup. In our summary discussions, we sometimes refer to the ``best execution time," which denotes the optimal performance achieved by a solver for a given problem across all tested core counts. This is a key part of the analysis, as parallel overhead can sometimes cause performance to degrade with a higher number of cores, and identifying this optimal point is important.
\end{enumerate}

\begin{figure}
    \centering
    \begin{subfigure}[b]{\columnwidth}
        \centering
        \includegraphics[width=0.7\columnwidth]{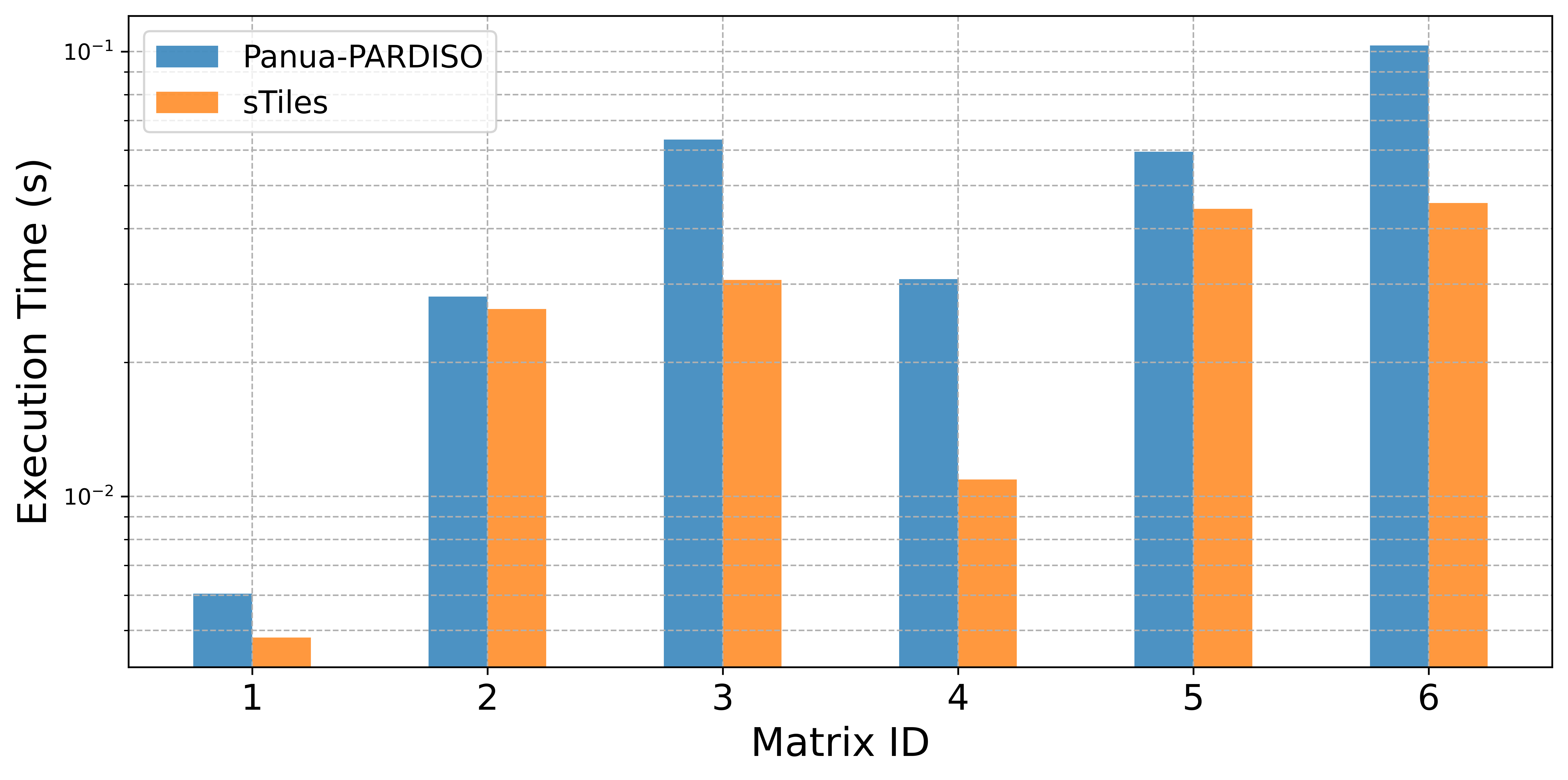}
    \end{subfigure}
    \begin{subfigure}[b]{\columnwidth}
        \centering
        \includegraphics[width=0.7\columnwidth]{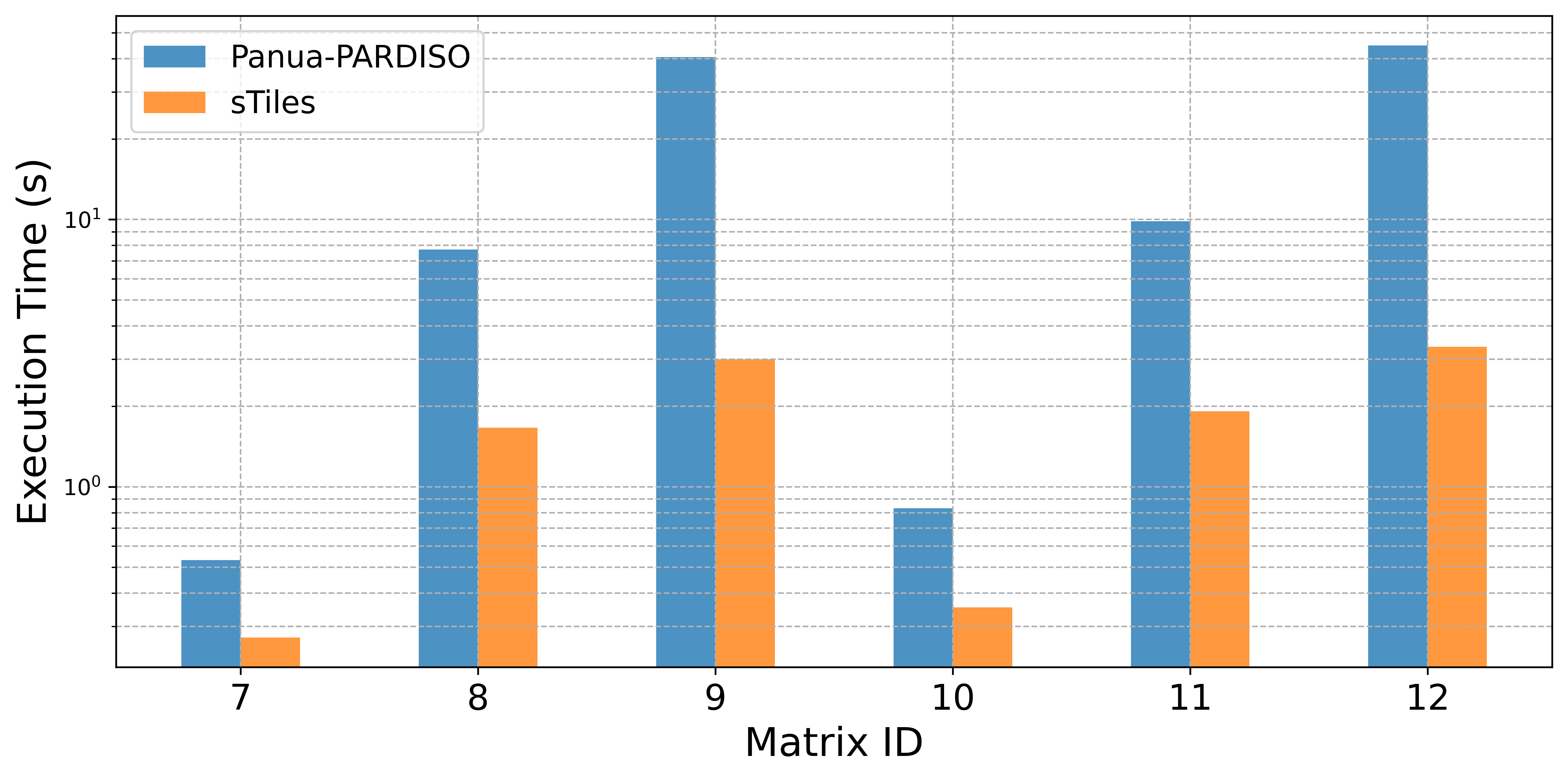}
    \end{subfigure}
    \begin{subfigure}[b]{\columnwidth}
        \centering
        \includegraphics[width=0.7\columnwidth]{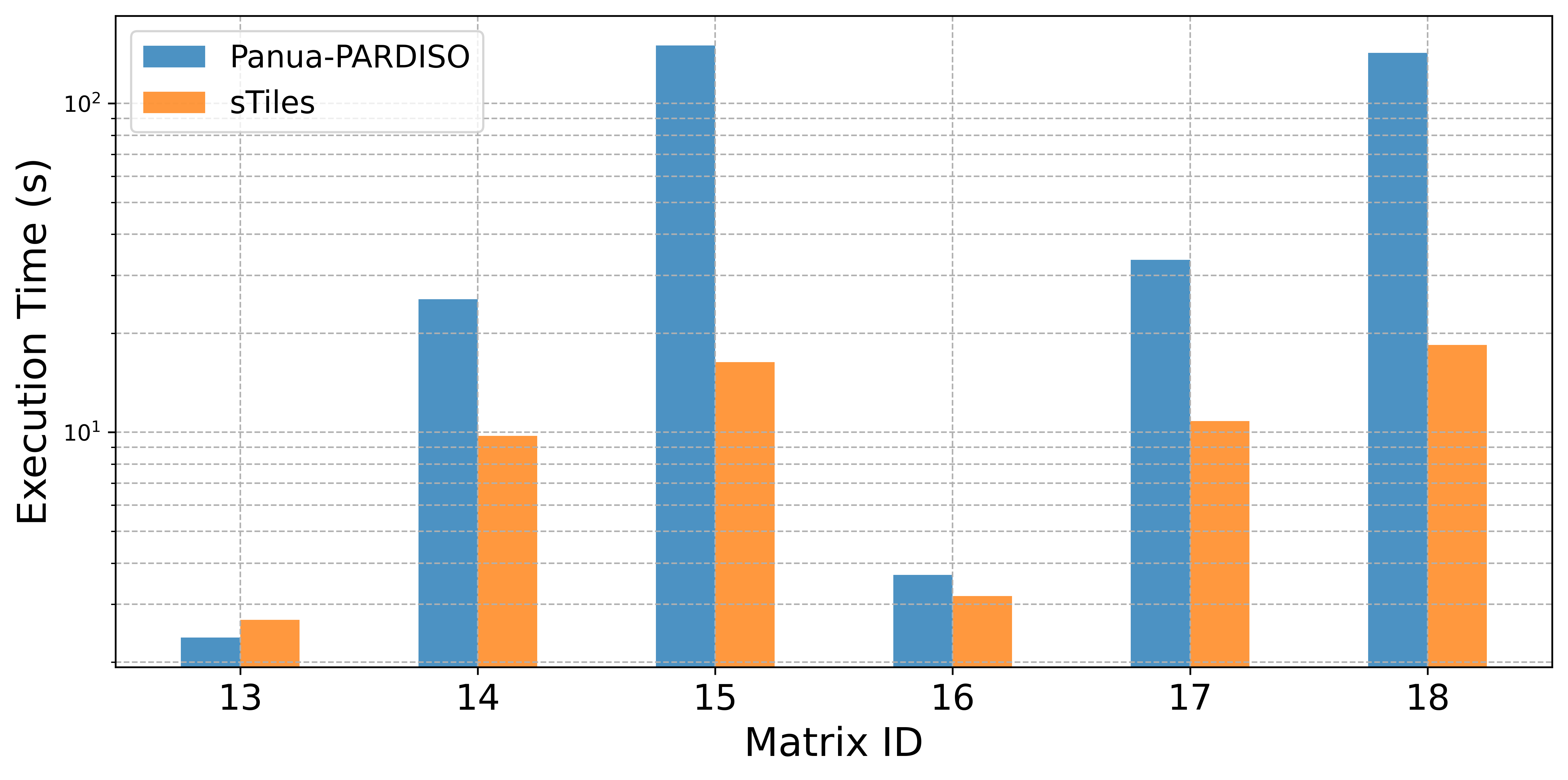}
    \end{subfigure}
    \caption{Performance comparison of \textit{sTiles} and Panua-PARDISO across different matrix configurations. Each subfigure represents a different matrix size category: small (top), medium (middle), and large (bottom).}
    \label{fig:results_set1}
\end{figure}
Figure~\ref{fig:results_set1} demonstrates the performance advantage of \textit{sTiles} over Panua-PARDISO across different matrix IDs ordered by increasing sizes. The performance gap becomes more pronounced as the matrix size increases (IDs 7-18), with \textit{sTiles} leveraging dense tiles to mitigate computational overhead and memory bandwidth limitations. In some cases, \textit{sTiles} achieves up to \textbf{13.49× speedup} compared to Panua-PARDISO.

\subsection{Parallel Scalability Evaluation}

\begin{figure*}[htbp]
    \centering
    \includegraphics[width=\textwidth]{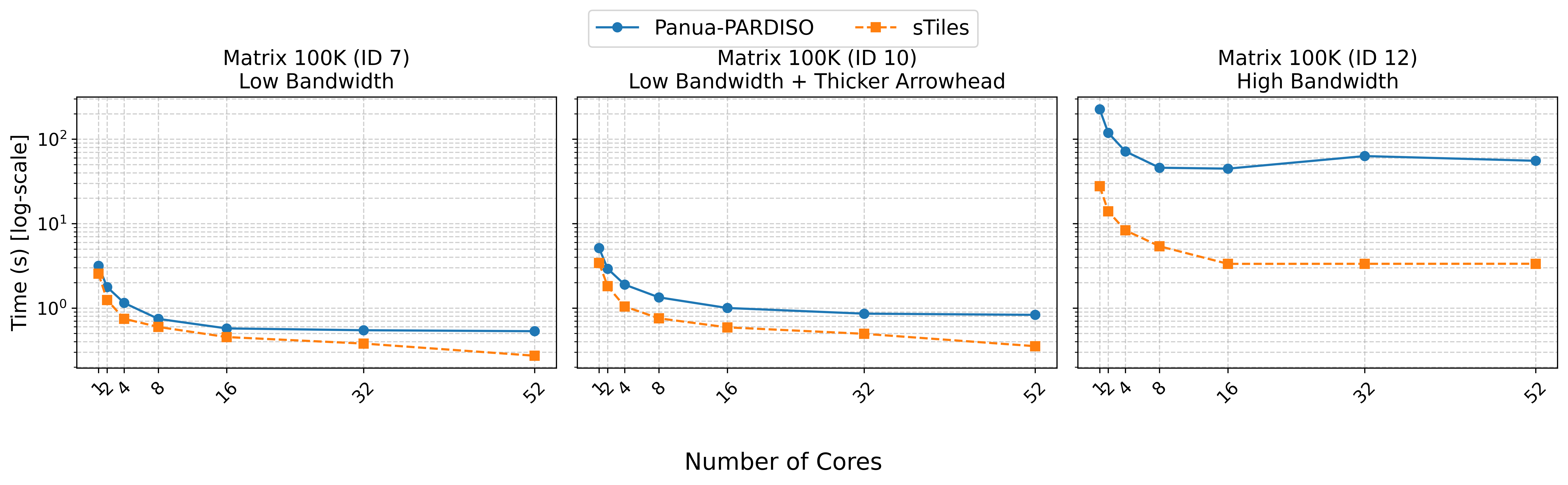}
    \caption{Scalability of \textit{sTiles} and Panua-PARDISO for selected matrix configurations. From left to right: a medium-sized matrix with low bandwidth (ID 7), a similar matrix with a thicker arrowhead structure (ID 10), and a matrix with high bandwidth and computational intensity (ID 12). \textit{sTiles} demonstrates superior scalability, particularly on larger and more computationally demanding problems.}
    \label{fig:scalability_comparison}
\end{figure*}

Figure~\ref{fig:scalability_comparison} presents the execution times of \textit{sTiles} and Panua-PARDISO across different core counts for two representative matrix sizes. For the smaller 10K matrix, both solvers exhibit strong parallel scalability. However, Panua-PARDISO consistently achieves shorter execution times in this scenario. The trend reverses for the larger 500K matrix, where \textit{sTiles} demonstrates superior parallel scalability and achieves increasingly lower execution times as the core count grows. This performance advantage highlights the efficiency of \textit{sTiles}'s parallelization strategy on large-scale problems.

\subsection{Impact of Sparsity on Performance}

The impact of sparsity on the computational performance of selected inversion is analyzed using a set of structured matrices with varying densities. Table \ref{tab:matrix_properties_stiles2} summarizes the properties of these matrices, including size, bandwidth, arrowhead thickness, and density. The density values, which exclude the arrowhead portion (see Figure \ref{fig:matrix_structure_illustration}), range from 0.010\% to 4.101\%, providing a broad spectrum of sparsity levels for evaluation.

\begin{figure}[!t]
    \centering
    \includegraphics[width=0.9\columnwidth]{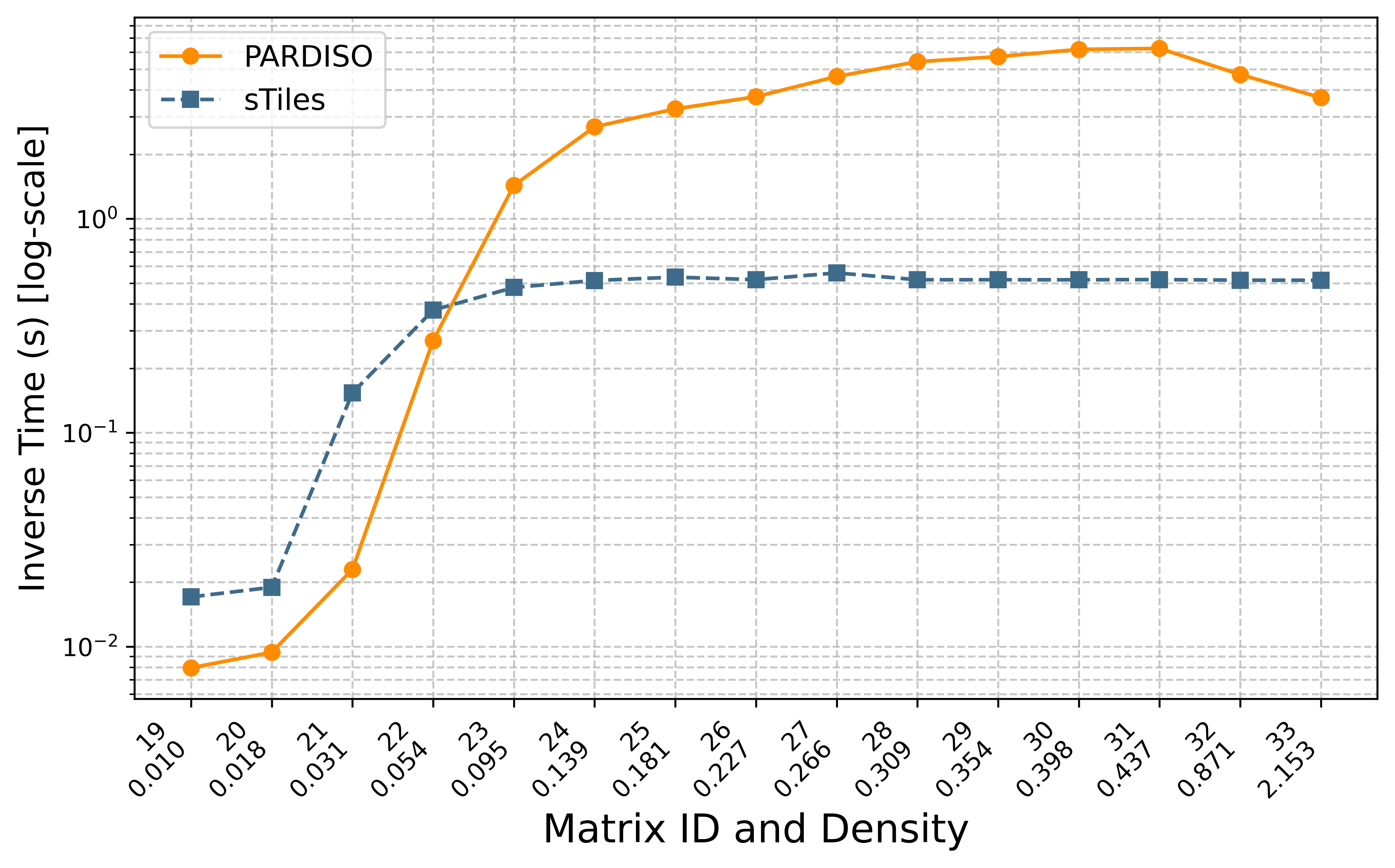}
    
    \vspace{1em}
    
    \includegraphics[width=0.9\columnwidth]{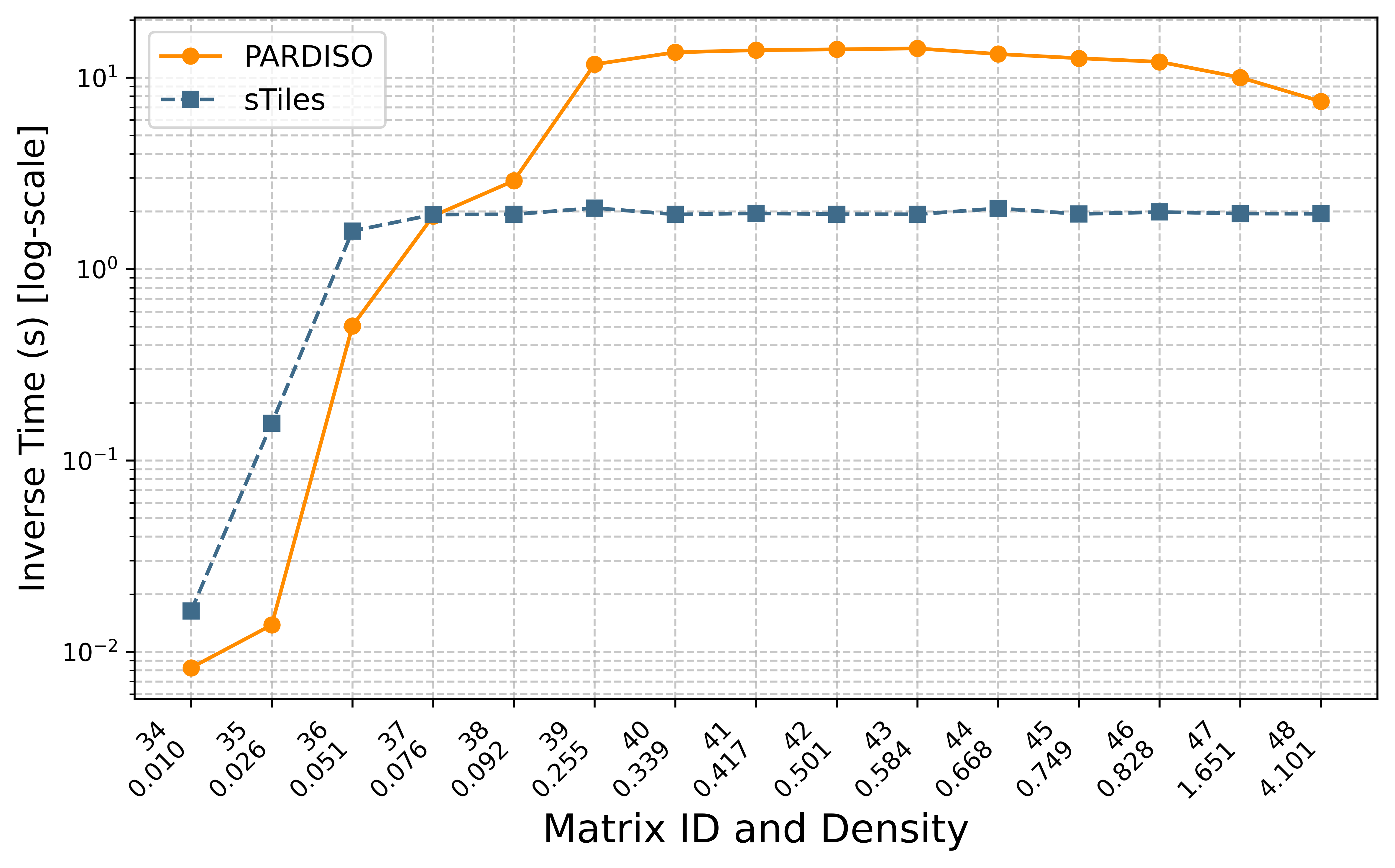}
    
    \caption{Selected inverse computation times of \textit{sTiles} and Panua-PARDISO across varying matrix densities. The top plot illustrates results for matrices with a bandwidth of 1500, while the bottom plot corresponds to matrices with a bandwidth of 3000. The x-axis represents matrix IDs, with density values displayed beneath each ID, while the y-axis uses a logarithmic scale to capture the variations in computation time.}
    \label{fig:results_set2}
\end{figure}

Figure \ref{fig:results_set2} presents the inverse computation times for both Panua-PARDISO and \textit{sTiles} solvers across matrices with increasing density. The results illustrate a clear distinction in solver behavior depending on the sparsity characteristics. For matrices with very low density (below 0.1\%), Panua-PARDISO exhibits a faster performance than \textit{sTiles}, primarily due to its optimized multifrontal structure for handling highly sparse matrices. However, as the density increases beyond this threshold, the performance of \textit{sTiles} stabilizes, while Panua-PARDISO experiences significant computational overhead.

The results indicate that \textit{sTiles} maintains consistent computational times across a wide range of density values, demonstrating its robustness for handling moderately sparse to dense matrices. In contrast, Panua-PARDISO's runtime increases substantially with density, reflecting the growing complexity of fill-in and symbolic factorization in direct solvers. For high-density matrices (greater than 1\%), \textit{sTiles} outperforms Panua-PARDISO, making it a more scalable approach for problems with increasing density.

This analysis highlights the advantage of the \textit{sTiles} approach in scenarios where matrix density increases while preserving a structured sparsity pattern. The ability of \textit{sTiles} to sustain lower computational cost across different density regimes makes it an attractive alternative for large-scale scientific computing applications where memory and time efficiency are critical constraints.

\begin{table}[!t]
  \caption{Matrix properties used in Cholesky factorization and selected inversion experiments for \textit{sTiles} – Set 2. Density values exclude the arrowhead part.}
  \label{tab:matrix_properties_stiles2}
  \centering
  \scriptsize
  \setlength{\tabcolsep}{0pt} 
  \begin{tabular*}{\columnwidth}{@{\extracolsep{\fill}} c c c c }
    \toprule
    \multicolumn{4}{c}{\makecell{\textbf{Matrix Size = 10,004} \\ \textbf{Arrowhead Thickness = 4}}} \\
    \addlinespace[4pt]
    \multicolumn{2}{c}{\textbf{Bandwidth = 1500}} & \multicolumn{2}{c}{\textbf{Bandwidth = 3000}} \\
    \cmidrule(r){1-2} \cmidrule(l){3-4}
    \textbf{ID} & \textbf{Density (\%)} & \textbf{ID} & \textbf{Density (\%)} \\
    \midrule
    19 & 0.010 & 34 & 0.010 \\
    20 & 0.018 & 35 & 0.026 \\
    21 & 0.031 & 36 & 0.051 \\
    22 & 0.054 & 37 & 0.076 \\
    23 & 0.095 & 38 & 0.092 \\
    24 & 0.139 & 39 & 0.255 \\
    25 & 0.181 & 40 & 0.339 \\
    26 & 0.227 & 41 & 0.417 \\
    27 & 0.266 & 42 & 0.501 \\
    28 & 0.309 & 43 & 0.584 \\
    29 & 0.354 & 44 & 0.668 \\
    30 & 0.398 & 45 & 0.749 \\
    31 & 0.437 & 46 & 0.828 \\
    32 & 0.871 & 47 & 1.651 \\
    33 & 2.153 & 48 & 4.101 \\
    \bottomrule
  \end{tabular*}
\end{table}

\subsection{Framework Robustness}
While our primary focus is on selected inversion, the underlying \emph{sTiles} framework is also highly efficient for full matrix inversion. To demonstrate this robustness, we evaluate the performance of \emph{sTiles} against PLASMA \cite{agullo2011towards, yarkhan2017porting}, a state-of-the-art numerical library, for full matrix inversion. The comparison highlights the fundamental difference in scheduling strategies: \emph{sTiles} employs a static scheduling approach designed to maximize data locality and minimize runtime overhead, whereas PLASMA utilizes a dynamic scheduler.

To ensure a fair comparison focused purely on the core computational kernels, our measurements for PLASMA exclusively time the inverse function. The initial data translation to a tile layout (\texttt{plasma\_omp\_dtr2desc}) and the final translation back to a standard LAPACK layout (\texttt{plasma\_omp\_ddesc2tr}) are excluded from the timing. This methodology isolates the performance of the inversion algorithm itself from data layout conversion costs.

\begin{figure}[htbp]
    \centering
    \includegraphics[width=0.85\columnwidth]{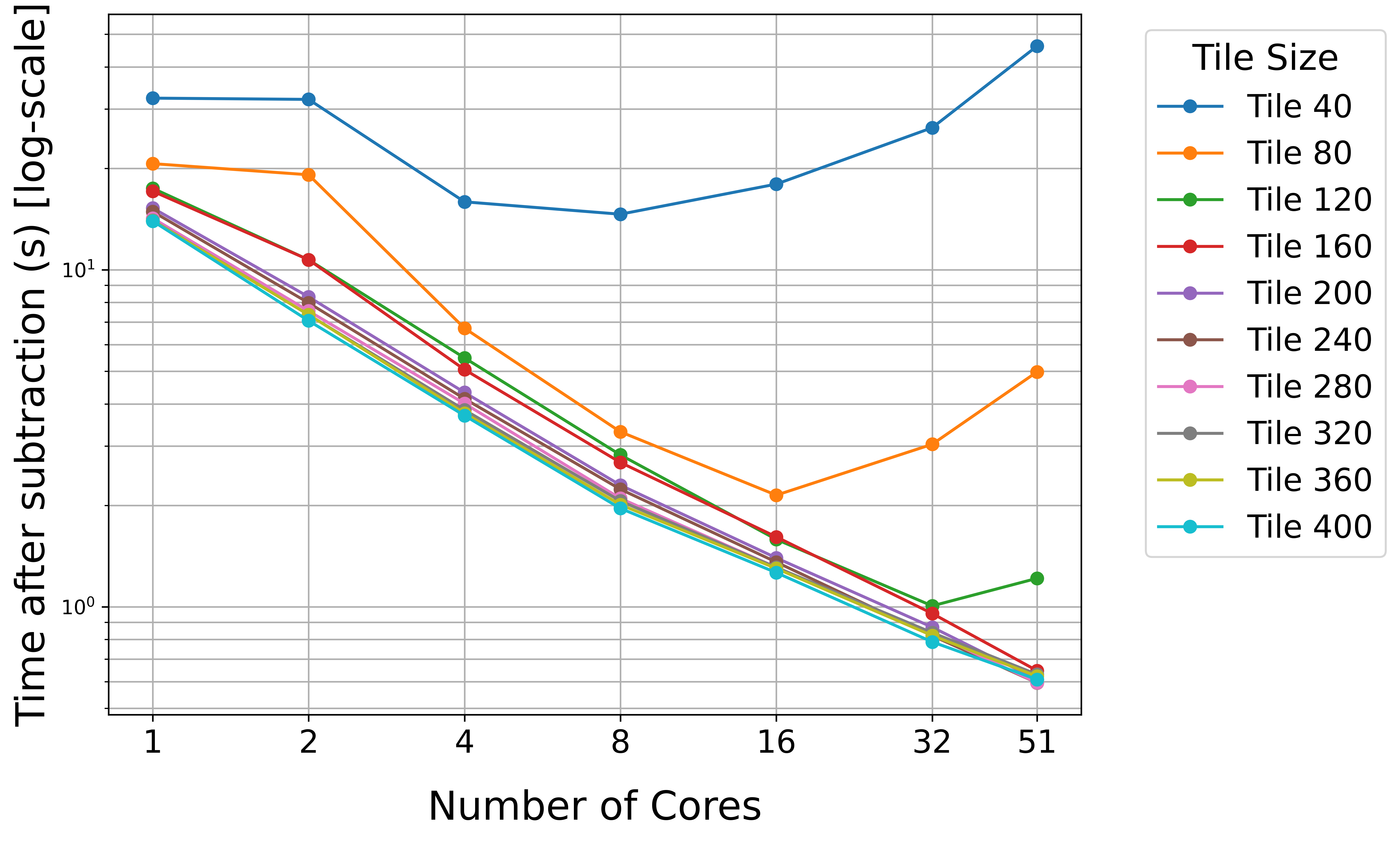}
    \caption{Execution times of PLASMA for full matrix inversion across different tile sizes and core counts for matrix ID 5.}
    \label{fig:full_dense_plasma}
\end{figure}

\begin{figure}[htbp]
    \centering
    \includegraphics[width=0.85\columnwidth]{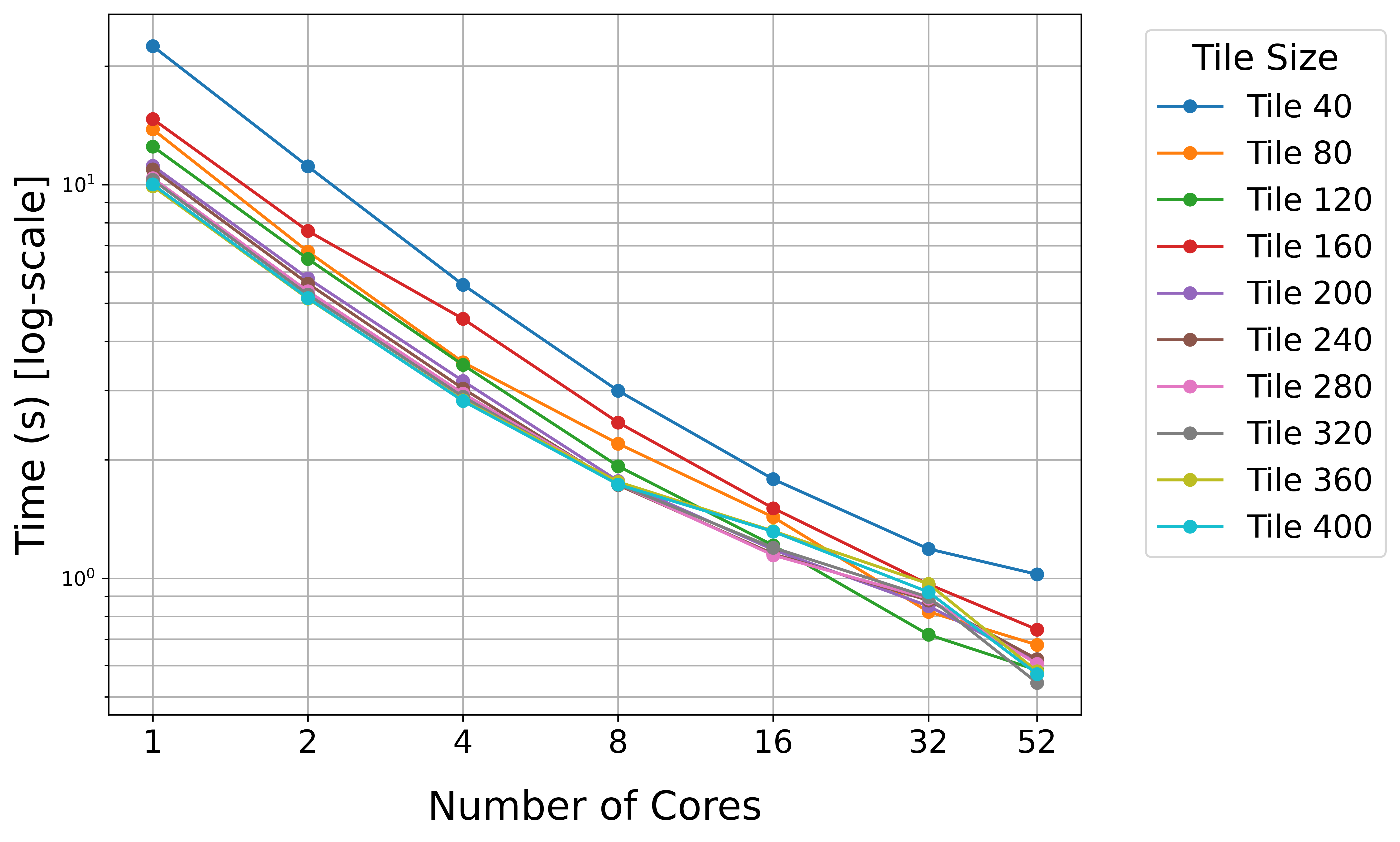}
    \caption{Execution times of \emph{sTiles} for full matrix inversion across different tile sizes and core counts for matrix ID 5.}
    \label{fig:full_dense_stiles}
\end{figure}

Figures \ref{fig:full_dense_plasma} and \ref{fig:full_dense_stiles} present the execution times for both libraries on matrix ID 5 (from Table \ref{tab:matrix_properties_stiles1}) across a range of tile sizes and core counts. The results reveal distinct performance characteristics:

\begin{itemize}
    \item \textbf{Scalability:} \emph{sTiles} demonstrates strong and consistent scalability, with execution time decreasing steadily up to 52 cores. In contrast, PLASMA's performance stagnates or degrades beyond 16 cores, particularly for smaller tile sizes (40 and 80). This suggests that the overhead from its dynamic scheduler and associated thread contention becomes a significant bottleneck at high core counts.
    \item \textbf{Tile Size Sensitivity:} \emph{sTiles} exhibits lower sensitivity to tile size, achieving robust performance across various configurations without requiring extensive tuning. PLASMA's performance is more variable with tile size, underscoring the challenge of managing its dynamic task dependencies effectively.
\end{itemize}

\begin{figure}[htbp]
    \centering
    \includegraphics[width=0.85\columnwidth]{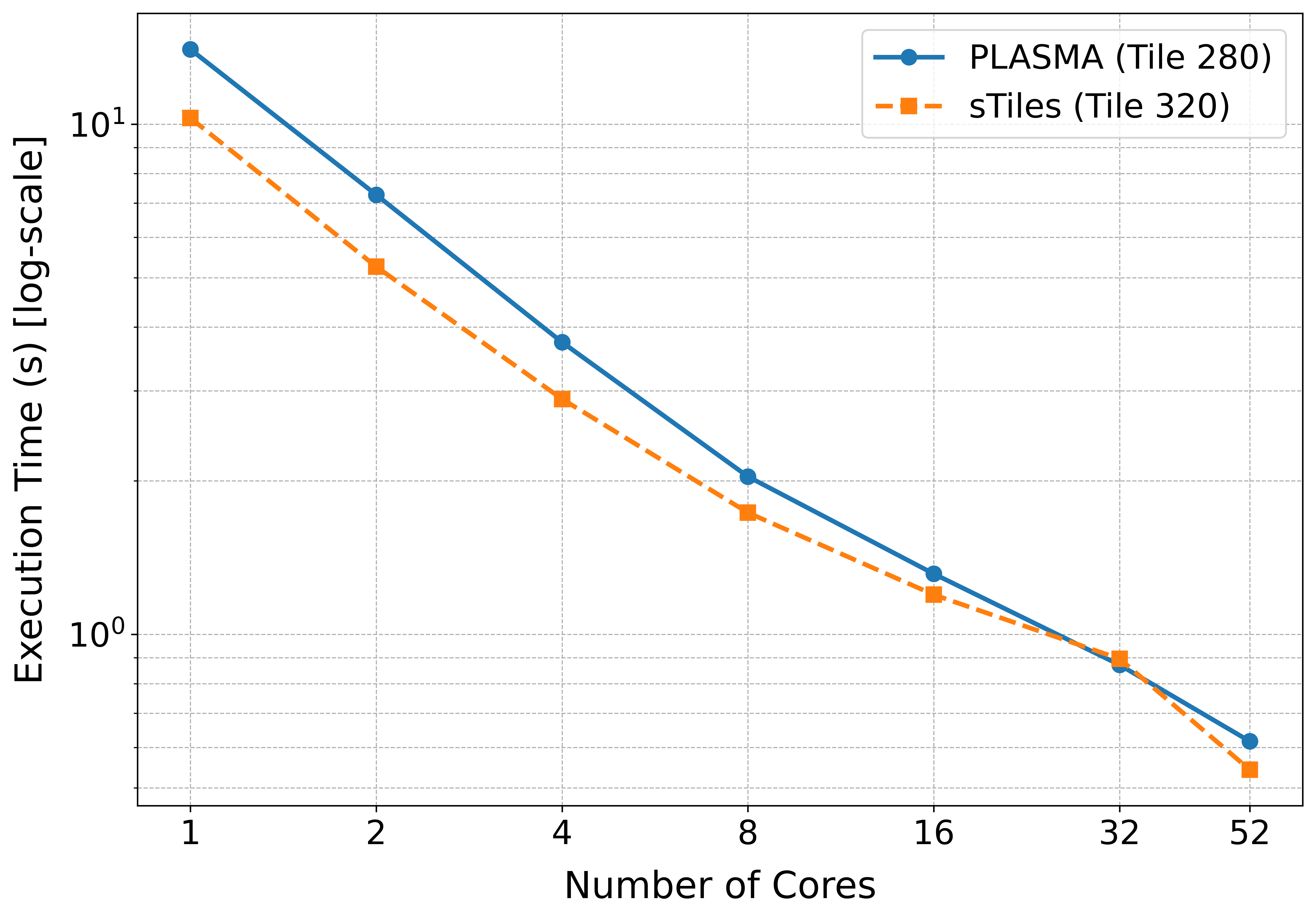}
    \caption{Full inversion comparison using the best tile size for each library (PLASMA: 280, \emph{sTiles}: 320). \emph{sTiles} demonstrates superior scalability at high core counts, outperforming PLASMA where parallel efficiency is most critical.}
    \label{fig:plasma_vs_stiles}
\end{figure}

To distill these findings, Figure \ref{fig:plasma_vs_stiles} provides a direct comparison using the optimal tile size for each library (280 for PLASMA and 320 for \emph{sTiles}). The results illustrate the classic trade-off between static and dynamic scheduling. \emph{sTiles} is significantly faster at lower core counts (1-16 cores) due to lower overhead. At 32 cores, a crossover occurs where PLASMA's dynamic scheduler achieves a marginal advantage, likely by better balancing a slightly uneven workload.

However, as the core count increases to 52, the low overhead of \emph{sTiles}'s static approach proves superior, and it reclaims the performance lead. The flattening curve for PLASMA indicates that the cost of dynamic task management becomes the primary bottleneck, limiting its scalability. This confirms that the static scheduling and data locality optimizations in \emph{sTiles} are more effective for achieving high parallel efficiency on many-core architectures.

\subsection{Accelerating Selected Inversion on GPU}

To evaluate the performance of our algorithm on massively parallel hardware, we leveraged a hybrid CPU-GPU implementation. In this model, CPU host threads orchestrate the task dependency graph and enqueue numerically intensive kernels onto an NVIDIA A100 GPU. Each thread manages a dedicated CUDA stream, maximizing concurrency. This subsection analyzes the resulting performance and explores the critical trade-offs involved in GPU acceleration.

\subsubsection{Execution Model and Performance Evaluation}
Before computation, the entire matrix is transferred to GPU memory. CPU threads then enqueue cuBLAS/cuSOLVER kernels onto their respective streams. Synchronization is handled on the CPU via the shared \texttt{core\_progress} array, with host threads waiting until dependencies are met before launching the next GPU kernel.

To demonstrate the potential of this approach, we selected a computationally demanding problem: the inversion of a large matrix (size 50,010, bandwidth 15,000) that has sufficient computational work to saturate the GPU and fits entirely within its 80 GB memory, thereby avoiding out-of-core complexities. For this direct comparison, all experiments were run on the \textbf{GPU Node} (AMD EPYC 7713 vs. NVIDIA A100).

A key factor in performance is the \textbf{tile size}, which was tuned for each architecture: a smaller size of \textbf{120} for the CPU to expose task parallelism across many cores, and a much larger size of \textbf{600} for the GPU to ensure that each kernel performs enough work to be efficient. This choice creates a fundamental trade-off: the large GPU tile size maximizes kernel throughput but reduces the overall number of parallel tasks available in the algorithm.

\begin{table}[htbp]
\centering
\caption{CPU vs GPU performance comparison for the two-phase selected inversion.}
\label{tab:cpu_gpu_comparison}
\scriptsize
\begin{tabular*}{\columnwidth}{@{\extracolsep{\fill}} c c c c }
\toprule
\makecell{\textbf{Cores/}\\\textbf{Streams}} & \makecell{\textbf{CPU Time}\\\textbf{(s)}} & \makecell{\textbf{GPU Time}\\\textbf{(s)}} & \makecell{\textbf{GPU Time}\\\textbf{+ Data (s)}} \\
\midrule
1   & 533.062 & 5.020 & 7.044 \\
2   & 282.714 & 2.517 & 4.524 \\
4   & 142.883 & 2.416 & 4.366 \\
8   & 75.225  & 2.015 & 3.954 \\
16  & 39.575  & 2.285 & 4.111 \\
32  & 26.861  & 1.913 & 3.562 \\
64  & 20.382  & 2.351 & 3.949 \\
\bottomrule
\end{tabular*}
\end{table}

As shown in Table~\ref{tab:cpu_gpu_comparison}, the GPU achieves its best time of 3.562 seconds with 32 streams, delivering a \textbf{5.72$\times$ speedup} over the best 64-core CPU time. The performance scales with the number of streams until the GPU is saturated with work, after which host-side synchronization overhead begins to limit further gains.

The observed speedup, while significant, is below the theoretical peak performance ratio of the hardware. This is an expected outcome due to three main factors:
\begin{enumerate}
    \item \textbf{Reduced Task Parallelism:} The large tile size required for GPU efficiency limits the number of concurrent tasks available in the algorithm's dependency graph.
    \item \textbf{Synchronization Overhead:} The dependency checks are managed by CPU threads, which introduces latency.
    \item \textbf{Data Transfer Cost:} The reported ``GPU Time + Data'' in Table~\ref{tab:cpu_gpu_comparison} includes only the \textit{final transfer of the inverted result} from device to host memory. 
    The initial matrix and Cholesky tiles are already on the GPU, and the inverse is initialized to zero directly in device memory, so no additional transfers occur during the selected inversion phase.

\end{enumerate}

In conclusion, our hybrid CPU-GPU model provides a substantial performance benefit for large-scale structured matrix inversion. The results validate that the \textit{sTiles} framework can effectively exploit GPU acceleration, with performance shaped by a clear trade-off between single-kernel efficiency (large tiles) and overall algorithmic parallelism (small tiles).

\section{Conclusion} \label{sec:conclusion}

In this work, we introduced an efficient GPU-accelerated parallel selected inversion algorithm for structured matrices using \emph{sTiles}. Our approach leverages a tile-based methodology that efficiently handles structured sparsity, particularly for arrowhead matrices, ensuring both computational efficiency and parallel scalability. By adopting a two-phase algorithm, we minimized interdependencies between computational tasks, allowing for efficient static scheduling and improved parallel execution across both CPU and GPU architectures.

The implications of this work extend beyond selected inversion, as the tile-based framework introduced in \emph{sTiles} can be extended to other numerical linear algebra problems that benefit from hybrid sparse-dense computations, offering a wide range of functionalities. Future work includes extending the framework to enabling multi-GPU support, supporting distributed-memory architectures, and optimizing for even larger-scale problems encountered in scientific computing, Bayesian inference, and high-dimensional statistical modeling. By continuously refining the \emph{sTiles} framework, we aim to push the computational boundaries of structured matrix computations and enhance its applicability in real-world high-performance computing applications.

\newpage
\appendix
\section{Algorithmic Complexity Derivation}
\label{app:complexity_derivation}

This appendix provides a detailed derivation of the computational complexity for the full and selected inversion algorithms discussed in Section~\ref{sec:complexity_main}. We consider a matrix of size \(n \times n\) partitioned into \(N \times N\) tiles of size \(b \times b\), such that \(n = N\,b\). The cost of a single tile operation (\textbf{GEMM}, \textbf{TRSM}, etc.) is \(O(b^3)\).

\subsection{Full Matrix Inversion}
The total work for inverting a dense matrix using our block algorithm is the sum of the costs of all tile operations. The number of calls for each dominant kernel is:
\begin{enumerate}
  \item \textbf{TRSM:} Called \(N\) times for the diagonal tiles. Total cost: \(W_{\text{TRSM}} = N \cdot O(b^{3})\).
  \item \textbf{LAUUM:} Called \(N\) times for the diagonal tiles. Total cost: \(W_{\text{LAUUM}} = N \cdot O(b^{3})\).
  \item \textbf{TRMM:} Called on all \(N(N-1)/2\) off-diagonal tiles. Total cost: \(W_{\text{TRMM}} = \frac{N(N - 1)}{2} \cdot O(b^{3})\).
  \item \textbf{GEMM:} This is the most computationally intensive part. The total number of \textbf{GEMM} calls is approximately \(\frac{N^3}{3}\). Total cost: \(W_{\text{GEMM}} = \frac{N^{3} - N}{3} \cdot O(b^{3}) = O(N^{3}b^{3})\).
\end{enumerate}
The overall complexity is dominated by the \textbf{GEMM} operations:
\[
W_{\text{Full}} = O(N^3 b^3) = O\bigl((Nb)^3\bigr) = O(n^3)
\]

\subsection{Selected Inversion (Arrowhead Band of Width \(B\))}
For the selected inversion of an arrowhead matrix, we only compute tiles within a band of width \(B\) tile blocks, plus the last tile in each row and column. This prunes a large number of \textbf{GEMM} operations. The total number of \textbf{GEMM} calls can be shown to be:
\begin{equation*}
\begin{split}
N_{\text{GEMM}} ={}& (N - B)B + \frac{B(B-1)}{2} \\
                   & + B^2(N - B - 1) + \frac{B(B+1)(2B+1)}{6}
\end{split}
\end{equation*}
To determine the asymptotic complexity for a large matrix where \(N \gg B\), we expand the expression and group terms by powers of \(N\):
\begin{equation*}
\begin{split}
N_{\text{GEMM}} ={}& (B^2 + B)N \\
                   & - \left(\frac{2B^3}{3} + \frac{B^2}{2} + \frac{B}{6}\right)
\end{split}
\end{equation*}
For \(B \ge 1\) and \(N \gg B\), the term linear in \(N\) dominates. The highest power of \(B\) in this dominant term is \(B^2\). Therefore, the number of \textbf{GEMM} calls is \(O(B^2 N)\).

The total work for the selected inversion is the number of \textbf{GEMM} calls multiplied by the cost per call:
\[
W_{\text{selected}} = O(B^2 N \cdot b^3)
\]
Substituting \(N = n/b\), we obtain the final complexity in terms of \(n\):
\[
W_{\text{selected}} = O\left(B^2 \frac{n}{b} b^3\right) = O(B^2 n b^2)
\]
This confirms that the method is significantly more efficient than full inversion when \(B \ll N\).

\end{document}